\documentclass[a4paper,twocolumn,11pt,accepted=2023-01-23]{quantumarticle}
\pdfoutput=1
\usepackage[utf8]{inputenc}
\usepackage[english]{babel}
\usepackage[T1]{fontenc}
\usepackage{amsmath}
\usepackage{hyperref}
\usepackage[dvipsnames]{xcolor}
\usepackage{amsfonts}
\usepackage{bbm}
\usepackage[utf8]{inputenc}
\usepackage[english]{babel}
\usepackage{subfig,dsfont}
\usepackage[T1]{fontenc}
\usepackage{amsmath}
\usepackage{hyperref}
\usepackage{tikz}
\usepackage[numbers]{natbib}
\usepackage{lipsum,ulem}

\newcommand\id{\mathbbm{1}}
\newcommand\ti{\text{i}}

\newcommand{\ket}[1]{\left|#1\right\rangle}

\hypersetup{
colorlinks = true,
linkcolor = Purple,
citecolor = Purple,
filecolor = Purple,
urlcolor = Purple,
}

\begin{document}
\title{Remotely Controlled Entanglement Generation}

\author{Ferran Riera-S\`abat}
\affiliation{Universit\"at Innsbruck, Institut f\"ur Theoretische Physik, Technikerstra{\ss}e 21a, 6020 Innsbruck, Austria}
\orcid{0000-0002-3038-5015}

\author{Pavel Sekatski}
\orcid{0000-0001-8455-020X}
\affiliation{University of Geneva, Department of Applied Physics, 1211 Geneva, Switzerland}

\author{Wolfgang D\"ur}
\affiliation{Universit\"at Innsbruck, Institut f\"ur Theoretische Physik, Technikerstra{\ss}e 21a, 6020 Innsbruck, Austria}
\orcid{0000-0002-0234-7425}

\maketitle

\begin{abstract}
We consider a system of multiple qubits without any quantum control. We show that one can mediate entanglement between different subsystems in a controlled way by adding a (locally) controlled auxiliary system of the same size that couples via an always-on, distant-dependent interaction to the system qubits. Solely by changing the internal state of the control system, one can selectively couple it to selected qubits, and ultimately generate different kinds of entanglement within the system. This provides an alternative way for quantum control and quantum gates that does not rely on the ability to switch interactions on and off at will and can serve as a locally controlled quantum switch where all entanglement patterns can be created. We demonstrate that such an approach also offers an increased error tolerance w.r.t. position fluctuations.
\end{abstract}

\section{Introduction}
\label{Sec.Introduction}

Quantum control is a crucial feature to generate entanglement and realize devices that can process quantum information. In a typical setup, local control of individual systems is assisted by tunable interactions or specific gates between selected qubits, that can be switched on and off at will. This allows one to entangle systems in a controlled way, and establish different kinds of entangled states that serve as a valuable resource for various tasks, ranging from quantum computation over quantum metrology to quantum networks. In turn, this demands advanced control of systems.

In many realistic scenarios, one does not have direct controllability and observability of the target system. One then considers the possibility to control such a system indirectly by bringing it in contact with an auxiliary system that one manipulates~\cite{weimer2010rydberg,albertini2018controllability,albertini2020subspace,albertini2021subspace,d2021dynamical}. Here, we consider the remotely controlled generation of entanglement and entangling gates in systems without quantum control and without tunable interactions. We show that apart from initialization in a homogeneous product state, no additional control on system qubits is required to produce different kinds of entangled states, including all graph states \cite{hein2006entanglement,briegelgraphstates}. We assume that a distant-dependent, always-on interaction couples the system qubits with an auxiliary control system of the same size. For commuting interactions --like in e.g., dipole-dipole interactions \cite{porras2004effective,pagano2020,joshi2020quantum}, and many other types of couplings-- we show that manipulating the internal state of the control system suffices to generate a large class of entangled states among the system qubits. The fixed, intrinsic interactions within the control system can be utilized to achieve this aim by only locally manipulating individual control qubits. The control system can be a selected sub-part of the total system we have access to, or a separate device that is, e.g., brought to a surface or a crystal structure and adds control to qubits within this system. The scheme works in case of always-on or no interactions among the system qubits themselves.

\begin{figure}[b]
    \centering
    \includegraphics[width=0.95\columnwidth]{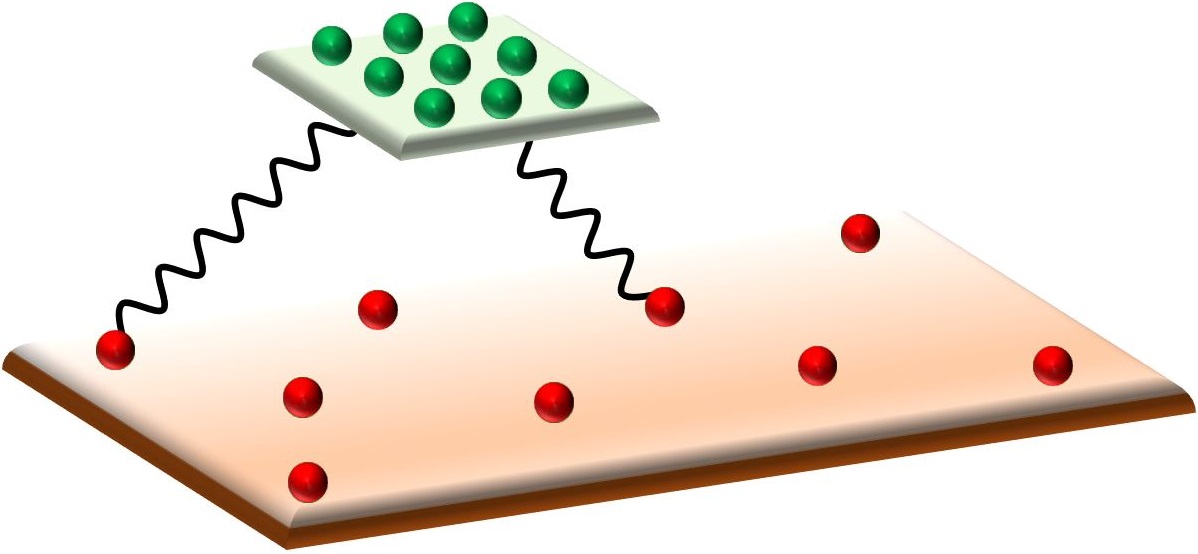}
    \caption{An auxiliary control system (green) couples selectively to a target system (red) and mediates an effective interaction by proper choice of the internal state of the control system.}
    \label{Fig1}
\end{figure}

The key element of our method is to use the spatial dependence of interactions, that induce different phases depending on the relative distance between the system and control qubits. By properly adjusting the internal state of the control device, one can make it selectively couple to several of the system qubits. This is done by using the multi-qubit control system as a single logical qubit, i.e., restricting its state to a two-dimensional subspace that is chosen in such a way that interactions with all but some selected system qubits yield a zero phase within this subspace. Such an approach is similar in spirit as used in the design of decoherence-free subspaces (DFS) \cite{sekatski2020optimal,W_lk_2020} and is applicable to all commuting interactions with spatial dependence, provided the size of the control system is sufficiently large. These selected pairwise interactions with the logical control system can be manipulated by different means, but importantly by only operating on the control system, to yield an effective interaction between selected qubits in the system, in such a way that the control system is factored out. In this sense, the control system serves to mediate interactions between the system qubits.

This article is organized as follows. In Sec.~\ref{Sec.Setting} we describe the underlying physical system and how to achieve a selective coupling of the control system to target qubits. In Sec.~\ref{Sec.Controlling.S} we show how we can control the target system, by preparing entangled states and by cancelling its inner interactions. In Sec.~\ref{Sec.Control.C} we discuss the different levels of control on the control system and we illustrate the full manipulation of the state of the control system by means of single physical qubit general operations. In Sec.~\ref{Sec.Resources.cost} we provide an analysis of the required resources for the setting in terms of single-qubit operations. In Sec.~\ref{Sec.Applications} we discuss different applications for the setting and we analyse two different physical examples based on trapped ions. In Sec.~\ref{Sec.Position.Noise} we analyse the effect of noise in the position of the target system and introduce techniques to reduce these effects. Finally, in Sec.~\ref{Sec.Conclusion} we conclude with an outlook and an overall discussion.

\section{Setting}
\label{Sec.Setting}

\subsection{Physical layer}

We consider an inaccessible target system of $n$ spatially distributed qubits, i.e., $S = \{ S_i \}_{i = 1}^n$ where qubit $S_i$ is located at position $\boldsymbol{r}_i^S$, and an additional fully controllable control system of $N$ qubits, $C = \{ C_i \}_{ i = 1 }^N$ where $C_i$ is located at $\boldsymbol{r}_i^C$, see Fig.~\ref{Fig1}. We also assume an inherent pairwise ZZ distance-dependent interaction, i.e., qubits $i$ and $j$ at positions $\boldsymbol{r}_i$ and $\boldsymbol{r}_j$, interact via $f_{ij} Z_i Z_j$ where $f_{ij} = J \, \text{f}\,(| \boldsymbol{r}_{i} - \boldsymbol{r}_{j} |)$ is the coupling strength that depends on the coupling constant $J$ and the distance between the two qubits. In particular, we consider $\text{f}(x) = x^{-1}$, but the setting is not restricted to this particular choice --- any other non-trivial distance dependence, e.g. of the form $x^{-\alpha}$ works as well.

The Hamiltonian describing the two systems is given by
\begin{equation*}
    H = H^C + H^S + H^{CS},
\end{equation*}
where
\begin{equation*}
\begin{aligned}
    & H^C = \sum_{1\leq i < j \leq N} f^C_{i j} \, Z^C_i Z^C_j \\
    & H^S = \sum_{1\leq i < j \leq n} f^S_{i j} \, Z^S_i Z^S_j \\
    & H^{CS} = \sum_{\substack{1\leq i \leq N \\ 1 \leq j \leq n}} f^{CS}_{i j} Z^C_i Z^S_j.
\end{aligned}
\end{equation*}
$H^C $ [$H^S $] describes interactions within the control [target] system, what we refer to as the \textit{self-interactions} of $C$ [$S$], and $H^{CS}$ describes interactions between qubits of different systems. In absence of self-interaction within $S$ the ZZ-coupling can be obtained from a general XYZ-interaction using fast control on $C$, as discussed in Appendix~\ref{app:heisemberg interaction}.

\subsection{Selective coupling to the target system}
\label{Sec.Selective.coupling}

Using the control system $C$ as a logical qubit, one can tune the interactions between $C$ and each of the qubits in $S$, to make it selective couple to any qubit of the target system $S$. This is achieved by restricting the state of the control system into a two-dimensional subspace of the form $span\{\ket{\boldsymbol{c}}, \ket{-\boldsymbol{c}}\}$, where $\ket{\boldsymbol{c}}$ is a state of the computational basis such that the components of vector $\boldsymbol{c} = (c_1, \dots, c_N)^T$ are given by $Z^C_i \ket{\boldsymbol{c} } = c_i \ket{\boldsymbol{c}}$. We consider non-integers values of $c_i$, as flipping the corresponding qubit at some intermediate time during the evolution allows one to obtain an arbitrary effective value $c_i \in [-1, 1]$. More details are given at the end of this section.

The interaction Hamiltonian between $C$ and qubit $S_j$ is diagonal in the computational basis, and if we define
\begin{equation}
\label{eq:lambdaj}
    \lambda_j \equiv \sum_{i=1}^N f^{CS}_{i j} c_i
\end{equation}
its eigenvalues in the logical subspace are given by $\{ \pm \lambda_j \}$, i.e.,
\begin{equation}
\label{eq:eXeXcom}
\begin{aligned}
    \sum_{i=1}^N f^{CS}_{i j} Z^C_i Z^S_j \ket{\pm\boldsymbol{c}}_C \ket{\bar{s}_j}_{S_j} & \\
    = \pm \lambda_j \, \bar{s}_j \ket{\pm\boldsymbol{c}}_C \ket{\bar{s}_j}_{S_j}&,
\end{aligned}
\end{equation}
where we refer to integer vector components as $\bar{s}_i \in \{-1,1\}$.

Therefore, when the control system is prepared in the logical qubit subspace $span\{\ket{\boldsymbol{c}}, \ket{-\boldsymbol{c}}\}$ the interaction Hamiltonian takes the form
\begin{equation*}
    H^{CS} = \sum_{j = 1}^n \lambda_j Z^C Z^S_j,
\end{equation*}
where $Z^C$ is the Pauli-Z operator acting on the logical subspace of $C$, i.e., $Z^C \ket{\pm \boldsymbol{c}} = \pm \ket{\pm \boldsymbol{c}}$, and $\lambda_j$ is the coupling strength between $C$ and qubit $S_j$.

The interaction pattern $\boldsymbol{\lambda} = ( \lambda_1, \dots, \lambda_n )^T$ only depends on the spatial distribution of the qubits and the vector $\boldsymbol{c}$ that we can freely choose by single qubit operations on $C$ (see below). In fact, if the system $C$ is large enough, i.e. contains enough qubits, any interaction pattern $\boldsymbol{\lambda}$ can be engineered by a proper choice of the logical subspace. Concretely, from Eq.~\eqref{eq:lambdaj} one can see that this can be done by setting the vector labelling the logical subspace to any vector $\boldsymbol{c}$ fulfilling
\begin{equation}
    \label{eq:FCL}
    \boldsymbol{F} \cdot \boldsymbol{c} = \boldsymbol{\lambda},
\end{equation}
where $\boldsymbol{F}$ is a $n\times N$ matrix given by $F_{ij} = f^{CS}_{ij}$. The matrix $\boldsymbol{F}$ only depends on the set of distances between qubits on $C$ and qubits on $S$, and hence, one can ensure that $\boldsymbol{F}$ is full-rank by properly arranging the qubits of the control system. Note that if $C$ is of the same size as $S$, i.e., if $N = n$, Eq.~\eqref{eq:FCL} can be inverted and hence any interaction pattern can be established. However, vector $\boldsymbol{c}$ has to be divided by $\text{max}_i |c_i|$ to ensure that $|c_k| \leq 1 \, \forall \, k$, what scales all interaction strengths by $\boldsymbol{\lambda} \to \boldsymbol{\lambda} / \text{max}_i |c_i|$. This implies for a given interaction pattern, there is a maximum coupling strength between $C$ and the qubits of $S$.

As we mentioned at the beginning of this section, we can effectively encode the logical qubit of $C$ in the subspace given by any vector $\boldsymbol{c}$ where $c_i \in [-1,1]$ by flipping the qubits during the evolution. If we initialize the control system in the subspace given by $\bar{\boldsymbol{c}} = (1,\dots, 1)$, the evolution between $C$ and $S$ at a time $\tau$ is given by $\exp\{ - \ti H^{CS} \tau \}$. Flipping qubit $C_i$ at time $t_i < \tau$ and at $\tau$ we obtain
\begin{equation*}
\begin{gathered}
    X^C_i e^{-\ti \sum_{k,l} f^{CS}_{kl} Z_k^C Z^S_l (\tau - t_i)} X^C_i e^{-\ti \sum_{k,l} f^{CS}_{kl} Z_k^C Z^S_l t_i} \\
    = e^{-\ti \sum_{k,l} f^{CS}_{kl} c_k Z_k^C Z^S_l \tau}
\end{gathered}
\end{equation*}
where $\boldsymbol{c} = (1,\dots, 1, c_i, 1\dots, 1)^T$ with $c_i = 2(t_i/\tau) - 1$, as the flip at $t_i$ turns $Z_i \to -Z_i$. Note that as
\begin{equation*}
    \left[ e^{-\ti H^{CS} t}, \, X^C_i e^{-\ti H^{CS} t'} X^C_i \right] = 0,
\end{equation*}
given any vector $\boldsymbol{c}$, we can implement the evolution in the effective logical subspace by flipping each qubit of $C$ at specific instants of the evolution, i.e., we apply $X^C_i$ at $t_i = (1 + c_i) \tau /2$ for $i=1,\cdots, N$. Observe, it is not necessary to flip again all the qubits at a time $\tau$, as it has no effect on the operation between $C$ and $S$. However, we need to take into account that at time $\tau$ the logical qubit in $C$ is now implemented in $\bar{\boldsymbol{c}} = (-1,\dots,-1)$. Therefore, the number of flips required to implement an effective logical subspace, $\eta_\lambda$, is upper-bounded by $\eta_\lambda \leq N$.

%----------------------------------------------------------------------------------------------------

\begin{figure}
\centering
\includegraphics[width=0.95\columnwidth]{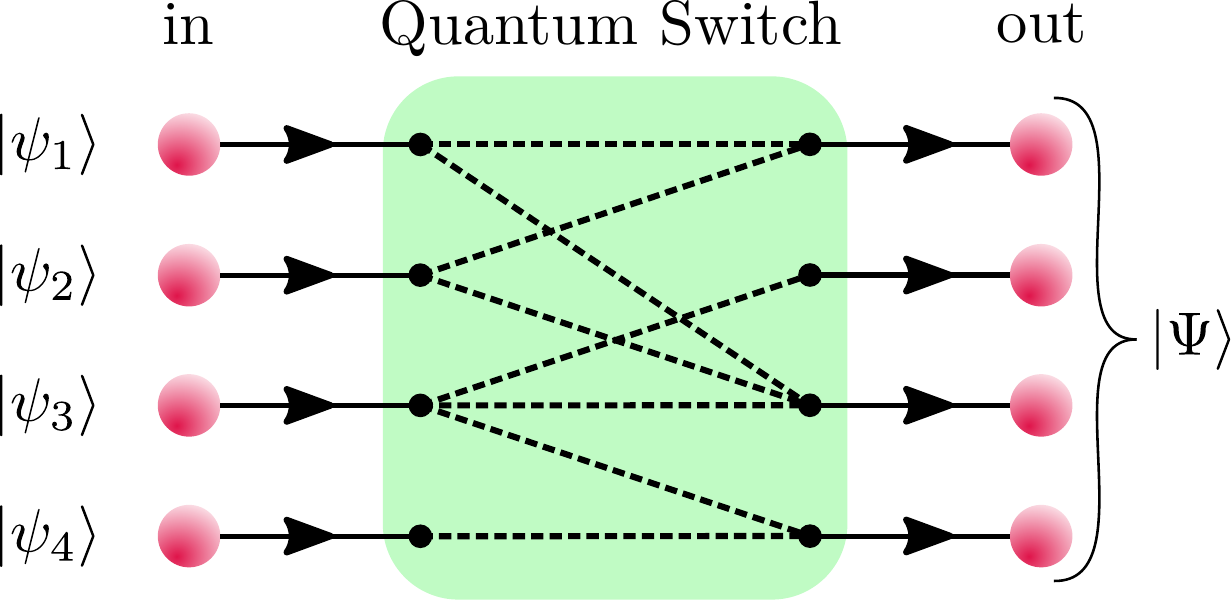}
\caption{Schematic representation of a quantum switch controlling the flow of information in a quantum network.}
\label{Fig:Switch}
\end{figure}

\section{Remote control of system \textit{S}}
\label{Sec.Controlling.S}

By choosing the logical subspace of the control system one can impose any interaction pattern between $S$ and $C$. This, combined with full control of $C$ that we assume, see Sec.~\ref{Sec.Control.C}, allows one to implement different effective interactions within $S$, which we can use to establish entangled states. The only requirement that we make on the target system is that it is initialized on a polarized state in the X-direction, i.e., in the state $\ket{+}^{\otimes n}$. We now illustrate different techniques to induce effective interactions and generate entanglement within $S$.

As it is shown later in Sec.~\ref{Sec.Cancel.selfint}, interactions within systems $C$ and $S$ do not affect our analysis, but for now, we ignore them, i.e., $H^C = H^S = 0$.

\subsection{Entanglement generation by gates sequence}
\label{Sec.Entanglement.generation.by.gate.sequence}

\begin{figure}
\centering
\vspace{-0.2in}
\subfloat[]{\includegraphics[width=0.4\columnwidth]{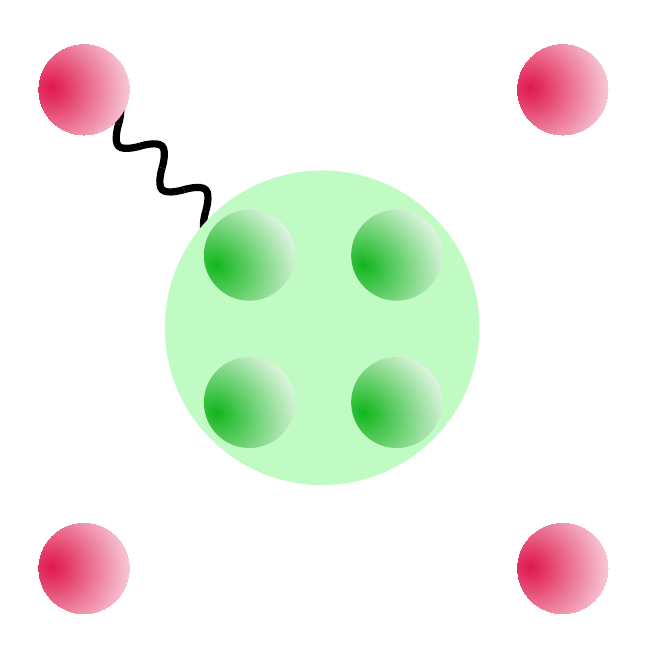}} \hspace{0.4in} \subfloat[]{\includegraphics[width=0.4\columnwidth]{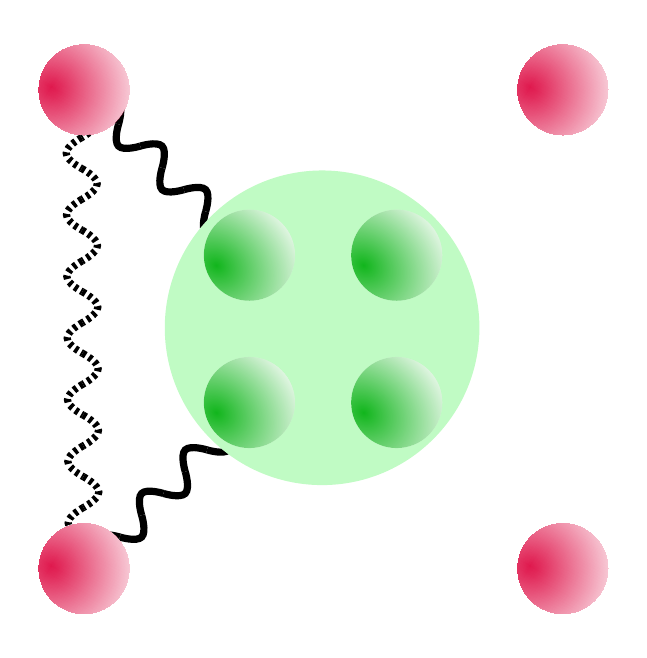}} \\
\subfloat[]{\includegraphics[width=0.4\columnwidth]{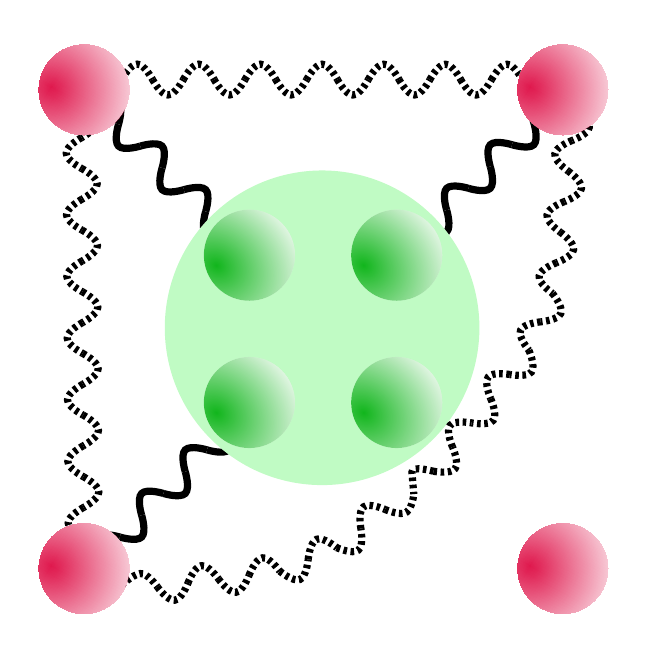}} \hspace{0.4in} \subfloat[]{\includegraphics[width=0.4\columnwidth]{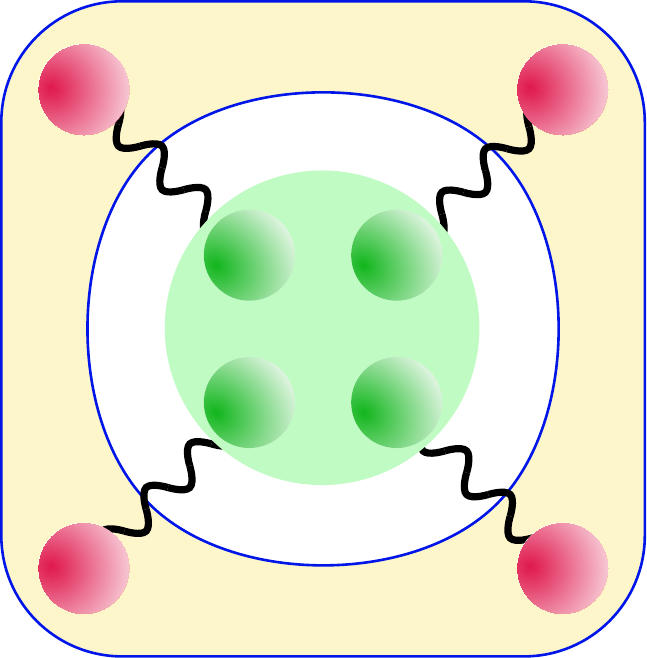}}
\caption{\label{Fig2D} Different operations that can be implemented in $S$ (external circles) by manipulating $C$ (inner circles). Solid wavy lines correspond to a coupling between a qubit and the control system, dashed wavy lines correspond to effective ZZ interactions and the shadow area corresponds to a multiple Z interaction. (a) Single qubit Z rotation. (b) Two qubits ZZ-interaction by coupling to two qubits plus an $X^C$-rotation. (c) Three qubits pairwise ZZ-interaction with the same method as in (b). (d) Multiple Z interaction by first entangling $C$ with the qubits followed by an X-rotation of $C$.}
\end{figure}

Without performing measurements, the effective interactions that we can implement within $S$ are of Z-type. As we demonstrate below, we can implement all diagonal unitary operations in $S$ by only controlling $C$, which suffices to generate any state of the form
\begin{equation}
\label{eq:allstates}
    \ket{\psi}_S = \frac{1}{ \sqrt{2^n} } \sum_{ \bar{s}_1, \dots, \bar{s}_n = 0}^1 e^{ \ti \theta_{1, \dots, n } } \ket{ \bar{s}_1, \dots , \bar{s}_n }_S,
\end{equation}
where $\bar{s}_i \in \{-1,1\}$. This includes the class of local maximally entangled (LME) states \cite{kruszynska2009local,KrausLMEprl,KrausLMEpra}, as well as all graph and hypergraph states. Graph states are a large class of highly entangled states \cite{hein2006entanglement,briegelgraphstates} that are a valuable resource in measurement-based quantum computation \cite{OneWayQcomputer, briegel2009measurement}, error correction \cite{nielsen2002quantum, campbell2008measurement, gottesman1997stabilizer,steane1998introduction} and quantum communication \cite{zwerger2016measurement}, including e.g., 2D cluster states and Greenberger–Horne–Zeilinger (GHZ) states. Any graph state can be written as
\begin{equation*}
    \ket{\Psi} = \prod_{(i,j) \in E} U_{ij} \ket{+}^{\otimes n},
\end{equation*}
where $E$ are the edges of the underlying graph and $U_{ij}$ are commuting, maximally entangling phase gates that we can generate here.

One method is to selective couple $C$ with a subset of $S$, i.e., $S' \subseteq S$, to implement
\begin{equation}
    \label{eq:US}
    U^{S'} = e^{-\ti \, \frac{\pi}{4} \sum_{j \in S'} Z^C Z^S_j },
\end{equation}
together with local operations on $C$, to generate an effective multi-qubit Z interaction on the selected subset $S'$. This is accomplished by applying the sequence
\begin{equation}
    \label{eq:method1}
    U^{S'} \, e^{-\ti \, \omega X^C } U^{S' \dagger} = e^{-\ti \, \omega G^C Z^{S'}_1 \! \cdots Z^{S'}_{n'} }
\end{equation}
and while leaving $C$ in an eigenstate of $G^C$, with
\begin{equation}
    \label{eq:G}
    G = \left( \frac{\ti}{2} \right)^{\! k} [ \cdots [ [ [ X, Z], Z ], Z ], \cdots Z],
\end{equation}
where we concatenate $k$ commutators. In this case, we need to take $k=n'$. Note that $G\in\{\pm X, \pm Y\}$ depending on the size of $S'$ given by $n'$. For $\omega = \pi / 4$, it generates a GHZ state in $S'$. Applying the same method on all subsets of qubits sequentially, and with a proper choice of induced interaction phases, one can generate any state of the form Eq.~\eqref{eq:allstates} (see Appendix~\ref{app:multi:Z}). For small interaction phases, this might however be costly, as maximally entangling gates between $C$ and $S$ are required to produce even a small interaction, and at least four maximally entangling gates are used to generate a maximally entangled state in $S$. This is however not an issue for graph states, where each edge corresponds to a $\pi/4$ interaction phase. In particular, the time required to implement Eq.~\eqref{eq:method1} is given by
\begin{equation}
    \label{eq:time.gates}
    t_g = 2 t_U + 2 t_V + t_{\text{i}} + \frac{\pi }{4 \omega},
\end{equation}
where $t_U$ is the time to implement $U^{S'}$ which depends on $S'$ and the spatial distribution of the qubits, $t_V$ is the time required to implement a logical Hadamard gate, and $t_{\text{i}} = t_V/2$ is the time to initialize the control system in the $\ket{+}_C$ state, see Sec.~\ref{Sec.Control.C}. Note that two Hadamard gates are required for the logical X rotation.

\subsection{Entanglement generation by projective measurements}
\label{Sec.Entanglement.generation.by.projective.measurements}

\subsubsection{Arbitrary graph states}

We can also prepare different states in the target system by first entangling it with the control system and then performing a projective measurement of the logical qubits in $C$, see Sec.~\ref{sec.logical.measurements}. In particular, by first entangling $C$ with $S'$ of size $n'$, and then measuring out the control system, we can directly implement a control-Z gate between all pairs of qubits in $S'$, i.e.,
\begin{equation}
\begin{aligned}
    \label{eq:CZij}
    R^{S'}_{\pm} \, \mathcal{M}_g^C \!\! : & \;\; U^{S'} \! \ket{+}_C \ket{\psi}_S \\ & \mapsto \prod_{(i<j) \in S'} \!\! \text{CZ}^S_{ij} \ket{\psi}_S
\end{aligned}
\end{equation}
where $\ket{\psi}$ is an arbitrary multipartite state, $U^{S'}$ is given in Eq.~\eqref{eq:US}, $\mathcal{M}^C_g$ is a projective measurement on the basis of $G$ given in Eq.~\eqref{eq:G} (with now $k = n'-1$) acting on $C$, and $\{R^{S'}_+=\id, R^{S'}_-= \bigotimes_{i\in S'} Z^S_i\}$ is a correction operation that depends on the outcome of $\mathcal{M}_g$. Note we can implement $R^{S'}_-$ by establishing the interaction pattern $\lambda_i = \lambda$ if $S_i\in S'$ and $\lambda_i = 0$ otherwise, and letting evolve $C$ and $S$ for a time $t = \pi/(2\lambda)$ while keeping the control system in the $\ket{0}_C$ state.

Repeating this method, we can prepare an arbitrary graph state in $S$ by sequentially creating each edge of the graph state. The time required to implement a control-$Z$ gate between two qubits on $S$ is given by
\begin{equation}
    \label{eq:time.measurements}
    t_m = t_U + t_\text{i} + t_\mathcal{M},
\end{equation}
where $t_\mathcal{M}$ is the time required to implement measurement $\mathcal{M}^C_g$. Therefore, the time to prepare an arbitrary graph state is given by $|E| \, t_m$, where $|E|$ is the number of edges.

\subsubsection{Decoherence-free GHZ states}

By performing projective measurements in the logical X basis on the control system, we also can prepare GHZ states in $S$ in the computation basis, i.e., of the form $(\ket{\bar{\boldsymbol{s}}} + \ket{-\bar{\boldsymbol{s}}})/\sqrt{2}$, up to random (but known) single-qubit X-gates. These states are of particular interest and once prepared they are not affected by the self-interaction in $S$, and hence they can be stored and used at later times without the need to cancel self-interaction resulting from $H^S$. In turn, states can only be manipulated further in a restricted way as some of the coherence among $S$ is destroyed.

There are different ways to obtain GHZ states on the computational basis. Here we introduce a general method that consists in iterating a Bell state generation procedure. The steps are the following:

\begin{enumerate}
\item Prepare a GHZ state between the logical qubit in $C$ and qubits $\{S_1, S_2 \}$, by applying $U^{S'}$. The final state is given by
\begin{equation*}
\begin{aligned}
    U^{S'} \ket{+}_C \ket{+\,+}_{S_1 S_2} & \\
    = \frac{1}{2} \Big[ \ket{-} \big(\ket{00}-\ket{11}\big) + \ti \ket{+} \big(\ket{01} + \ket{10}\big) & \Big].
\end{aligned}
\end{equation*}

\item Perform a projective measurement $\mathcal{M}_x^C$ to the logical qubit in $C$. The state of $S'$ collapses to
\begin{equation*}
\begin{aligned}
    R^{S_1}_{\pm} & \mathcal{M}_x^C \!\! : \, U^{S'} \ket{+}_C \ket{+\,+}_{S_1 S_2} \\
    &\mapsto \frac{1}{\sqrt{2}} \ket{\pm}_C \left( \, \ket{\bar{\boldsymbol{s}}_\pm}_{S_1S_2}+\ket{ -\bar{\boldsymbol{s}}_\pm}_{S_1S_2} \right),
\end{aligned}
\end{equation*}
where $\bar{\boldsymbol{s}}_{\pm} = (1,\pm1)$. Note that using the control system we can tune the relative phase in the state but not change vector $\bar{\boldsymbol{s}}$ which depends on the measurement outcome.

\item Add qubit $S_3$ to the entangled state by first applying $U^{S''}$ where $S'' = \{S_1, S_3\}$, i.e.,
\begin{equation*}
\begin{aligned}
    U^{S''} \frac{1}{\sqrt{2}} \ket{+}_C \big(\ket{\bar{\boldsymbol{s}}}_{S'} +& \ket{-\bar{\boldsymbol{s}}}_{S'} \big) \ket{+}_{S_3} \\
    = \frac{1}{2} \Big[ \ket{-} \big( \ket{\bar{\boldsymbol{s}},0} - & \ket{-\bar{\boldsymbol{s}},1} \big) \\
    + \ti \ket{+} \big( & \ket{\bar{\boldsymbol{s}},1} + \ket{-\bar{\boldsymbol{s}},0} \big) \Big],
\end{aligned}
\end{equation*}
and next performing measurement $\mathcal{M}^C_x$ followed by $R^{S_1}_\pm$. Qubits $S_1$, $S_2$ and $S_3$ end up in a state of the from $(\ket{\bar{\boldsymbol{s}}'} + \ket{-\bar{\boldsymbol{s}}'})/\sqrt{2}$ where $\bar{\boldsymbol{s}}'$ depends on the two measurement outcomes.

\item Note step 3 is independent of the size of $S'$ and therefore it can be iterated to sequentially add more qubits to the entangled state and prepare an arbitrarily large GHZ state.
\end{enumerate}

\subsection{Entanglement generation by control rotation}
\label{Sec.Entanglement.generation.by.control.rotation}

A third method for implementing an effective ZZ-interaction between two qubits consists in simultaneously coupling $C$ to the two qubits, and applying an additional field in the X-direction on the logical qubit in $C$. At a certain time $\tau$, the state of $C$ factors out, while an effective ZZ-interaction between the two qubits in $S$ is generated. The evolution generated by
\begin{equation*}
    H_x = \lambda Z^C \left( Z^S_1 + Z^S_2 \right) + \omega X^C,
\end{equation*}
of duration $\tau$  results in the global unitary transformation
\begin{align*}
    e^{-\ti \, \tau H_x}&= \cos \! \left(\tau \sqrt{\lambda^2 \left( Z_1^S+Z_2^S\right)^2+\omega^2} \, \right) \mathds{1}^C \\
    &- \ti \, \frac{\sin \! \left(\tau \sqrt{\lambda^2\left( Z_1^S+Z_2^S\right)^2+\omega^2} \, \right)}{\sqrt{\lambda^2\left( Z_1^S+Z_2^S\right)^2+\omega^2}} H_x.
\end{align*}
The eigenvalues of $Z_{1}^S+Z_2^S$ are $2,0$ and $-2$. For any state $\ket{\psi}_{S_1 S_2}$ of the target system with $\big(Z_{1}^S+Z_2^S\big) \ket{\psi}_{S_1 S_2} = 0$, we have $e^{- \ti H_x \tau}\ket{+}_C \! \ket{\psi}_{S_1 S_2}  = e^{-\ti \, \omega \tau}\ket{+}_C \ket{\psi}_{S_1 S_2}$ since the control qubit is prepared in an eigenstate of $X^C$. For a state $\ket{\psi}_{S_1 S_2} $ with $\big(Z_{1}^S+Z_2^S\big) \ket{\psi}_{S_1 S_2} = \pm 2\ket{\psi}_{S_1 S_2}$, we tune the interaction time to $\tau = \pi / \sqrt{\omega^2 + 4 \lambda^2}$ guaranteeing that $\sin\big(\tau \sqrt{\lambda^2(\pm 2)^2+\omega^2}\,\big)=0$ and $e^{- \ti H_x \tau}\ket{+}_C \! \ket{\psi}_{S_1 S_2}  =-\ket{+}_C \ket{\psi}_{S_1 S_2}$. Combining the two observations one concludes that the resulting evolution reads
\begin{equation}
\label{eq:evolutionX}
    e^{-\ti H_x  \tau} \! \ket{+}_C \! \ket{\psi}_{S_1 S_2} \!\! = \! e^{-\ti \phi Z^S_1 Z^S_2} \! \ket{+}_C \! \ket{\psi}_{S_1 S_2}\!,
\end{equation}
where $\ket{\psi}$ is an arbitrary two-qubits state and $\phi = \frac{\pi}{2}\left( 1 - \omega / \sqrt{\omega^2 + 4 \lambda^2} \, \right)$ is an arbitrary phase.
% and by tuning $\lambda$ and $\omega$ (and choosing the required interaction time $\tau$) one can set any effective coupling between the qubits, i.e.,
% \begin{equation}
% \label{eq:evolutionX}
%     e^{-\ti H_x  \tau} \! \ket{+}_C \! \ket{\psi}_{S_1 S_2} \!\! = \! e^{-\ti \phi Z^S_1 Z^S_2} \! \ket{+}_C \! \ket{\psi}_{S_1 S_2}\!,
% \end{equation}
% where $\ket{\psi}$ is an arbitrary two-qubits state, $\tau = \pi / \sqrt{\omega^2 + 4 \lambda^2}$ and $\phi = \frac{\pi}{2}( 1 - \omega / \sqrt{\omega^2 + 4 \lambda^2} \, )$. 
That is, for any desired interaction phase $\phi$, one can find $\lambda,\omega$ and the corresponding time $\tau$, to generate an effective interaction phase between the two target qubits in $S$, while keeping the control system decoupled.

The same method allows one to simultaneously generate a maximally entangling pairwise interaction between three qubits, i.e., to implement $\exp\{-\ti \, \frac{\pi}{4} \sum_{i,j=1}^3 Z^S_i Z^S_j \}$. For this we couple $C$ equally to the three qubits, set $\omega = \lambda \sqrt{5}/\sqrt{3}$, and the state is prepared at time $\tau = \sqrt{3} \pi / (2 \lambda \sqrt{2} )$ where $\tau \sqrt{\lambda^2(\pm 1)^2+\omega^2} =\pi$ and $\tau \sqrt{\lambda^2(\pm 9)^2+\omega^2} =2 \pi$. See Appendix~\ref{app:gate:maentic:field} for details.

Note in Eq.~\eqref{eq:evolutionX} we cannot implement an arbitrary interaction pattern by sequentially applying the evolution in integer subspaces as explained in Sec.~\ref{Sec.Selective.coupling}, as the evolution generated in this case does not fulfil Eq.~\eqref{eq:eXeXcom}. Therefore, the evolution should be implemented employing Hamiltonian simulation techniques, i.e., we use the Trotter formula to reproduce the desired evolution
\begin{equation*}
    \lim_{k \to \infty} \left[ e^{-\ti \lambda Z^C \big( Z^S_1 + Z^S_2 \big) \tau/k} e^{-\ti \omega X^C \tau/k}\right]^k = e^{-\ti H_x  \tau}.
\end{equation*}
When the time evolution is split in such a way, the $\exp\{-\ti \lambda Z^C \big(Z^S_1 + Z^S_2 \big)\}$ interaction can be implemented by flipping the qubits of the control system as described in Sec.~\ref{Sec.Resources.cost}. This shows that the method can be implemented even if only local control on the $C$ is available. In this case, however, it is highly demanding in terms of number of operation cost. In contrast, it is efficient in scenarios where direct global control on $C$ is available.

More methods to generate entanglement are detailed in Appendix~\ref{app:gate:sequence} and~\ref{app:Hamiltonian:simulation}. This includes sequences of three gates to directly establish a Bell pair or other techniques from Hamiltonian simulation where the alternating application of non-commuting pairwise interactions between $C$ and two different qubits leads to a three-qubit interaction in a second order.

\subsection{Cancellation of self-interactions}
\label{Sec.Cancel.selfint}

So far we have ignored interactions between qubits of the same ensemble, i.e., $H^C = H^S = 0$. Even though self-interactions are present in our physical model, they do not affect our analysis. $H^S $ commutes with $H^{CS}$, which means at any time, one can use $C$ to apply a correction operation that generates an effective pairwise ZZ interaction in $S$ that cancels all inherent self-interactions. Note that in principle $H^S$ can be stronger than the coupling $H^{CS}$ in which case self-interactions cannot be cancelled dynamically. However, for a fixed time the phases induced by self-interactions need only to be cancelled modulo $2 \pi$, this can always be done for a long enough time. One may also use the self-interactions to speed up the entanglement generation between qubits, rather than cancelling it. Hence, in the case of self-interactions, the desired entanglement state is established at a fixed time only, and further operations are required to maintain it.

On the other hand, in $C$, we do not need any correction operation since it is not affected by its self-interactions. Restricting its state into a logical sub-space of the form $span \{ \ket{\boldsymbol{c}} , \ket{-\boldsymbol{c}} \}$ we are also making it insensitive to $H^C $, i.e., any pair of states $\ket{\boldsymbol{c}}$ and $\ket{-\boldsymbol{c}}$ have the same eigenvalue for $H^C $. Therefore, within the subspace, the evolution generated by $H^C $ just yields a global phase.

%----------------------------------------------------------------------------------------------------

\section{Control of the logical system \textit{C}}
\label{Sec.Control.C}

One of the requirements of our setting is to have full control of the logical qubit in $C$. In this section, we discuss the scenario where full control is directly assumed, and then we also show that actually, control of its individual physical qubits suffices to achieve a full manipulation of the logical qubit (unitaries and measurements).

\subsection{Global control}
\label{Sec.Global.Control}

We first consider the situation where the control system $C$ is given by a small quantum processor of $N$ qubits, for which full control is ideally assumed and any desired operation on the control system can be realized by a sequence of elementary single- and two-qubit gates. Notice that the restriction to a particular two-dimensional subspace as outlined above actually limits the required operations. Under this assumption, effective spin values can be obtained, and any of the methods to remotely generate entanglement between the system qubits described above is applicable, including the one described in Sec.~\ref{Sec.Entanglement.generation.by.control.rotation}. That is, our methods allow one to extend the control of the small-scale quantum processor to remote systems.

\subsection{Local control}
\label{Sec.Local.Control}

In the following, we show how we can implement any single logical qubit gate and arbitrary projective logical measurements by using the inherent physical qubit-qubit interaction and applying single physical qubit operations on the control system.

\subsubsection{Logical single qubit gates}

Some gates on the logical subspace can be decomposed as a combination of single physical qubit gates. For instance, an arbitrary Z-rotation in the Bloch sphere, $R_z(\phi) = e^{-\ti \phi Z/2}$, can be applied to the logical qubit by individually rotating the qubits of $C$, i.e.,
\begin{equation*}
    R^C_z (\varphi) = R^C_{z,1} (\varphi_1) \otimes \cdots \otimes R^C_{z,N} (\varphi_N),
\end{equation*}
where $\varphi = \sum_i \varphi_i / c_i$. Notice that an operation on a single physical qubit also suffices to obtain the desired logical Z-rotation.

However, in general, a single qubit gate on the logical qubit is an entangling gate between the physical qubits, and hence it requires entangling physical qubit operations to be implemented. Therefore, we generally use the self-interaction term with local control of the physical qubits to control the logical qubit. In particular, a single qubit gate, $V$, on the logical qubit can be implemented as
\begin{equation}
\label{eq:VL}
    V^C = \Bigg( \prod_{i=2}^N \text{CX}^C_{1 \to i} \Bigg) \, V^C_1 \, \Bigg( \prod_{j=2}^N \text{CX}^C_{1 \to j} \Bigg).
\end{equation}
As we explain below, by performing fast flips of the physical qubits in the control system, we can isolate the interaction between any pair of physical qubits in $C$. This combined with single physical qubit gates, we can implement a control-X gate between any pair of physical qubits in $C$. In this way, single qubit control suffices to implement any gate on the logical qubit.

In the protocols shown in Sec.~\ref{Sec.Controlling.S} we need the control system in the $\ket{+}_C$ state. In this case, we can initialize the control system by first measuring each physical qubit in the $Z$ basis to obtain the $\ket{0}_C$, and then apply a Hadamard gate to the logical qubit using Eq.~\eqref{eq:VL}, as $\text{H}_{\text{ad}} \ket{k} = (\ket{0}+(-1)^k\ket{1})/\sqrt{2}$. Note this procedure would require $2N-2$ physical qubit control gates. However, as the control system is known to be in the $\ket{\boldsymbol{0}}_C$ state, we also can prepare the $\ket{+}_C$ state by first encoding the logical qubit in the subspace given by $\bar{\boldsymbol{c}}=\boldsymbol{1}=(1,\dots,1)$ and then applying
\begin{equation}
\label{eq:prep}
   \left[ \prod_{i=2}^N \text{CX}^C_{1\to i}\right]  \text{H}_{\text{ad},1}^C \ket{\boldsymbol{1}}_C = \frac{\ket{\boldsymbol{1}} + \ket{-\boldsymbol{1}}}{\sqrt{2}}= \ket{+}_C.
\end{equation}
Note that this initialization procedure only requires $N-1$ physical qubit control gates, and hence, it is twice faster to implement than the logical Hadamard gate. Finally, the effective logical subspace is established by performing flips during the interaction as described in Sec.~\ref{Sec.Selective.coupling}.

In order to make use of the self-interactions in the control system, $H^C$, while implementing a control gate between a pair of physical qubits in $C$, the control qubits must leave the logical qubit subspace which is designed to be insensitive to $H^C$. During this process, in order to avoid the control qubits interacting with the target system and between themselves in an uncontrolled way, we set an effective null spin value for all pairwise interactions within $C$ and interactions with external systems by flipping them at certain times.

We show how we can ``turn off'' interactions within $C$ by detailing the procedure. First, we divide the ensemble into two subsets of the same size $\Omega_0$ and $\Omega_1$, e.g., $\Omega_0 = \{C_1,\dots, C_{N/2}\}$. Note, we can cancel out all interactions between the two sets at time $2t$ by flipping the qubits in $\Omega_0$ at $t$, i.e.,
\begin{equation*}
\begin{gathered}
    W^{(1)} \equiv X^C_1 \cdots X^C_{N/2} \, e^{-\ti H^C t} X^C_1 \cdots X^C_{N/2} \, e^{-\ti H^C t} \\
    = e^{-\ti \big( \sum_{k < l \leq \frac{N}{2}} f^C_{kl} Z^C_k Z^C_l + \sum_{\frac{N}{2} < k < l} f^C_{kl} Z^C_k Z^C_l \big) \, 2t}.
\end{gathered}
\end{equation*}
As $W^{(1)}$ only contains interaction within $\Omega_0$ and within $\Omega_1$, we split $\Omega_0 = \Omega_{00} \cup \Omega_{01}$ ($\Omega_1 = \Omega_{10} \cup \Omega_{11}$). We can iterate the previous step but now flipping the qubits in $\Omega_{00} \cup \Omega_{10}$ what cancels out half of the interactions in $W^{(1)}$, i.e.,
\begin{equation}
\label{eq:w2}
    W^{(2)} \equiv \! \Bigg[ \prod_{i \in \Omega_{00}\cup\Omega_{10}} \!\!\! X^C_i \Bigg] W^{(1)} \Bigg[ \prod_{j \in \Omega_{00}\cup\Omega_{10}} \!\!\! X^C_j \Bigg] W^{(1)}.
\end{equation}
We proceed by iterating this step, i.e., applying twice the previous evolution $W^{(j)}$ with an X flip to half of the qubits of each previous subset between the concatenation. In the $k = \left \lceil \log_2 N \right \rceil$ step each subgroup $\Omega$ contains a single qubit and, hence, all interactions are cancelled, i.e., $W^{(k)} = \id$. Note that $k$ steps correspond to concatenate $2^k (\leq 2N)$ times the gate $\exp\{-\ti H^C t \}$, where between each application of the evolution we perform $N$ flips. However, as one can see in Eq.~\eqref{eq:w2}, half of the flips overlap with the next step, and we do not have to count them. Also, we have to add an extra simultaneous flip to all of the qubits of $C$ at $2^k t/2$ to cancel out interactions of $C$ with $S$. Observe the global flip does not affect the interactions within $C$. Therefore, the total number of flips to ``turn off'' the interactions of $C$ is given by $N + 2^{\left \lceil \log_2 N \right \rceil-1} N$.

Slightly adapting the procedure just described, we can isolate the interaction between any pair of qubits by flipping the two qubits simultaneously between each concatenation. This keeps cancelling the interactions with the rest of the qubits but leaves a clean interaction between the pair. The interaction plus an extra single-qubit operation allows us to implement a control-X gate. Therefore, as in Equations~\eqref{eq:VL} and \eqref{eq:prep}, $N-1$ and $2N-2$ control gates are performed respectively, the number of flips required to initialize the control system, $\eta_{\text{i}}$, and to implement a general single logical qubit gate, $\eta_V$, are bounded by
\begin{equation*}
\begin{aligned}
    & \eta_{\textit{i}} \leq N (N^2-1) \\
    & \eta_V \leq 2N (N^2-1)
\end{aligned}
\end{equation*}
where we used that $2^{\left \lceil \log_2 N \right \rceil-1} \leq N$. This implies that the required spin flips in system $C$ to perform arbitrary logical operations scale as $O(N^3)$, i.e. polynomial in the number of qubits.

\subsubsection{Logical projective measurements}
\label{sec.logical.measurements}

Any projective measurement of the logical qubit can be effectively performed by individually measuring each of the physical qubits in $C$. As shown in \cite{walgate2000local}, one can always distinguish between two orthogonal states by local operations. Given two $N$-qubit orthogonal states $\ket{\psi}$ and $\ket{\psi^\perp}$ we can write them as
\begin{equation*}
\begin{gathered}
    \big|\psi\big\rangle_C = \big|u\big\rangle_{C_1} \big|v\big\rangle_{C_2 \dots C_N} + \big|u^\perp\big\rangle_{C_1} \big|w\big\rangle_{C_2 \dots C_N} \\
    \big|\psi^\perp\big\rangle_C = \big|u\big\rangle_{C_1} \big|v^\perp\big\rangle_{C_2 \dots C_N} + \big|u^\perp\big\rangle_{C_1} \big|w^\perp\big\rangle_{C_2 \dots C_N}
\end{gathered}
\end{equation*}
where $\{ \ket{u}, \big|u^\perp\big\rangle \}$ is an orthonormal basis, $\{ \ket{v}, \big|v^\perp \big\rangle \}$ and $\{ \ket{w}, \big|w^\perp \big\rangle \}$ are two pairs of (non-normalized) orthogonal states. After measuring qubit $C_1$ in the basis $\{\ket{u}, \big|u^\perp\big\rangle \}$, the problem reduces to distinguishing between the orthogonal states $\ket{v}$ and $\big|v^\perp\big\rangle$ (or $\ket{w}$ and $\big|w^\perp\big\rangle$) of the remaining $N-1$ qubits. After repeating the same procedure to all the parts, one distinguishes between the two original states.

Note that this measurement does not project the state of the control system into $\ket{\psi}$ or $\big|\psi^\perp\big\rangle$ but in some other (known) state. However, if $C$ is entangled with $S$, we can collapse the state of $S$ in the same way as the actual projective logical measurement, i.e., if $\mathcal{M}_\Psi$ is a concatenation of single physical measurements that distinguishes between $\ket{\psi}$ and $\big|\psi^\perp\big\rangle$ then
\begin{equation*}
\begin{aligned}
    \mathcal{M}_\Psi^C \! : & \; \gamma_\alpha \ket{\psi}_C \ket{\alpha}_S + \gamma_\beta \big|\psi^\perp\big\rangle_C \ket{\beta}_S \\
    \mapsto &
    \begin{cases}
    \; \big| \tilde{\psi} \big\rangle_C \ket{\alpha}_S & \text{with prob. } p = |\gamma_\alpha|^2 \vspace{0.03in} \\
    \; \big|\tilde{\psi}^\perp\big\rangle_C \ket{\beta}_S & \text{with prob. } 1-p = |\gamma_\beta|^2,
    \end{cases}
\end{aligned}
\end{equation*}
where $\ket{\alpha}$ and $\ket{\beta}$ are two arbitrary $n$-qubit states and $\big|\tilde{\psi}\big\rangle$ and $\big|\tilde{\psi}^\perp\big\rangle$ are two random $N$-qubit product states that depend on the outcome of the $N$ single-qubit measurements performed to implement $\mathcal{M}_\Psi$. The choice of $\ket{\psi}$ and $\ket{\psi^\perp}$ within the logical subspace hence allows one to effectively perform the corresponding two-outcome measurement, i.e., an arbitrary projective measurement on the logical qubit.

%----------------------------------------------------------------------------------------------------

\section{Resources cost}
\label{Sec.Resources.cost}

Here we analyze the resources in terms of the number of flips, $\eta$, required for our setting. We call $\eta_\lambda$ the number of flips required to implement the evolutions in different effective logical subspaces, and $\eta_l$ the number of flips required in the control of the logical qubit, i.e., $\eta = \eta_l + \eta_\lambda$. Recall that we denote by $n$ the number of target qubits in $S$ and by $N$ the number of qubits of the control system $C$.

\textit{Entanglement generation by gates sequence.} In Sec.~\ref{Sec.Entanglement.generation.by.gate.sequence} and in Appendix~\ref{app:multi:Z}, we show how we can prepare any state of the form Eq.~\eqref{eq:allstates} by applying at most $2^n-1$ gate sequences of the form Eq.~\eqref{eq:method1}. Therefore, as for each subset $S'$ we need to apply $U^{S'}$ and $U^{S' \dagger}$, and in Sec.~\ref{Sec.Controlling.S}, we showed that for implementing $U^{S'}$ at most $N$ flips are required, the total number of flips to implement the target interaction fulfils $\eta_\lambda \leq 2 N (2^n-1)$. On the other hand, $\eta_l \leq 2 \eta_V (2^n-1) + \eta_{\text{i}}$, as first we need to initialize the control system in the $\ket{+}_C$ state and two Hadamard gates are needed to implement $\exp\{-\ti \omega X^C\}$ what is used in each gate sequence. Therefore, this method requires at most $\mathcal{O}(2^n N^3)$ spin flips. We remark that for the preparation of graph states, at most $n(n-1)/2$ phase gates are required. Hence, in this case, the total number of flips is reduced to $\mathcal{O}(n^2 N^3)$.

\textit{Entanglement generation by projective measurements.} To analyse the method detailed in Sec.~\ref{Sec.Entanglement.generation.by.projective.measurements}, we consider the preparation of an arbitrary graph state $\ket{G}$ with $|E|$ edges and $V$ vertices, see Eq.~\eqref{eq:CZij}. We showed for each edge we need to implement one gate of the form $U^{S'}$, and hence $\eta_\lambda \leq |E| N$. Also in this case we need to prepare the control system to the $\ket{+}_C$ state $E$ times, as after each measurement the state of the control system is collapsed, and hence $\eta_l = |E| \eta_{\text{i}}$. Therefore, this method requires at most $\mathcal{O}(|E| N^3)$ spin flips. Again, since the number of edges is upper bounded by $n(n-1)/2$, the required flips are polynomial in $n$ and $N$.

\textit{Cancellation of self-interactions.} To cancel the effects of $H^C$ at a specific time, we need to implement an interaction between each pair. The method introduced in Sec.~\ref{Sec.Entanglement.generation.by.gate.sequence} corresponds to sequentially implementing an effective interaction between each of the $n(n-1)/2$ pairs of qubits in $S$. Therefore, if we use the method introduced in Sec.~\ref{Sec.Entanglement.generation.by.gate.sequence}, $\eta_\lambda \leq N n(n-1)$ and $\eta_l = \eta_V n(n-1) + \eta_{\text{i}}$. The total number of flips is given by $\mathcal{O}(n^2 N^3)$.

%----------------------------------------------------------------------------------------------------

\section{Applications and examples}
\label{Sec.Applications}

The method we present above is completely general and can be utilized in different setups. One can also envision applications for different quantum processing tasks.

\subsection{Applications}

In the context of quantum computation, our approach can be used as a novel design principle to realise a quantum processor in a setup where local control is available, while interactions between qubits are not tunable and cannot be manipulated selectively. Half of the qubits serve as a control system to mediate interactions and realize two and multi-qubit gates between the remaining qubits that form the quantum register. One can envision that the whole system consists of the same type of qubits, but it is also conceivable that the control system and the quantum register correspond to different types, e.g., a well-controllable NV center with some surrounding nuclear spins, or a system of a few trapped ions that is brought close to a crystal or a surface to control interactions within atoms there. Notice that a self-interaction among the system qubits, i.e., the quantum register, is not required - only that the qubits that couple to the control system. Here we have shown that controlling the internal state of $C$ suffices to mediate any diagonal unitary operations on $S$. Notice that our technique can also be combined with a more standard approach of controlling the coupling $H^{CS}$ directly, e.g., by moving the control system. The combination of the two control techniques might allow to increase the coupling strengths, or reduce the size of $C$.

For quantum networks, a crucial element is a quantum switch \cite{vardoyan2019stochastic, coopmans2021netsquid} that allows one to selectively transmit quantum information between input and output ports, Fig.~\ref{Fig:Switch}. Our approach provides this functionality in an entanglement-based way, where any desired entanglement pattern (that can be used to transmit quantum information via teleportation between the ports) can be generated.

\subsection{Examples: Trapped ions}
\label{Sec.Examples}

\begin{figure*}
\centering
\subfloat[]{\includegraphics[width=0.62\columnwidth]{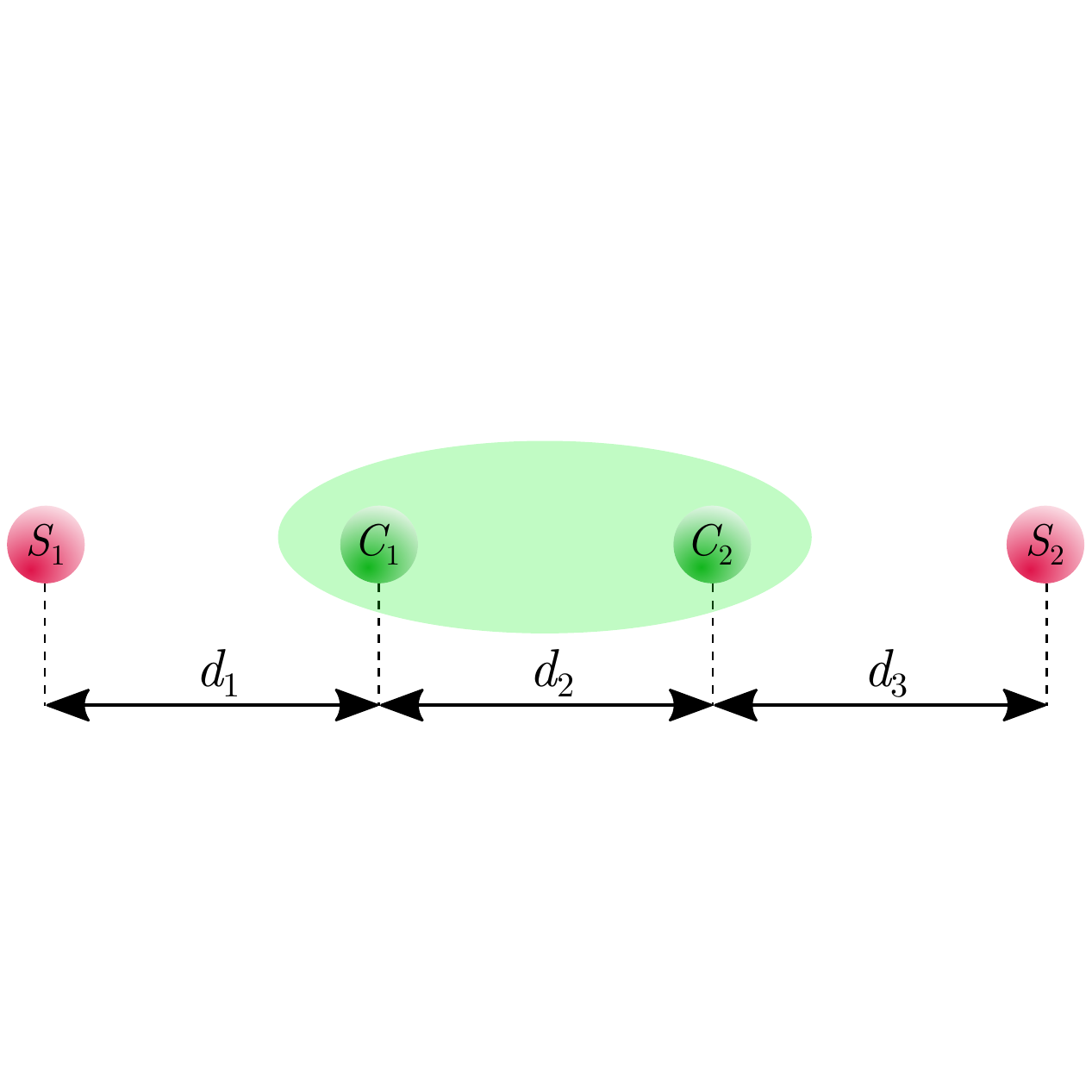} \label{fig:SM:1D}} \hspace{0.5cm}
\subfloat[]{\includegraphics[width=0.62\columnwidth]{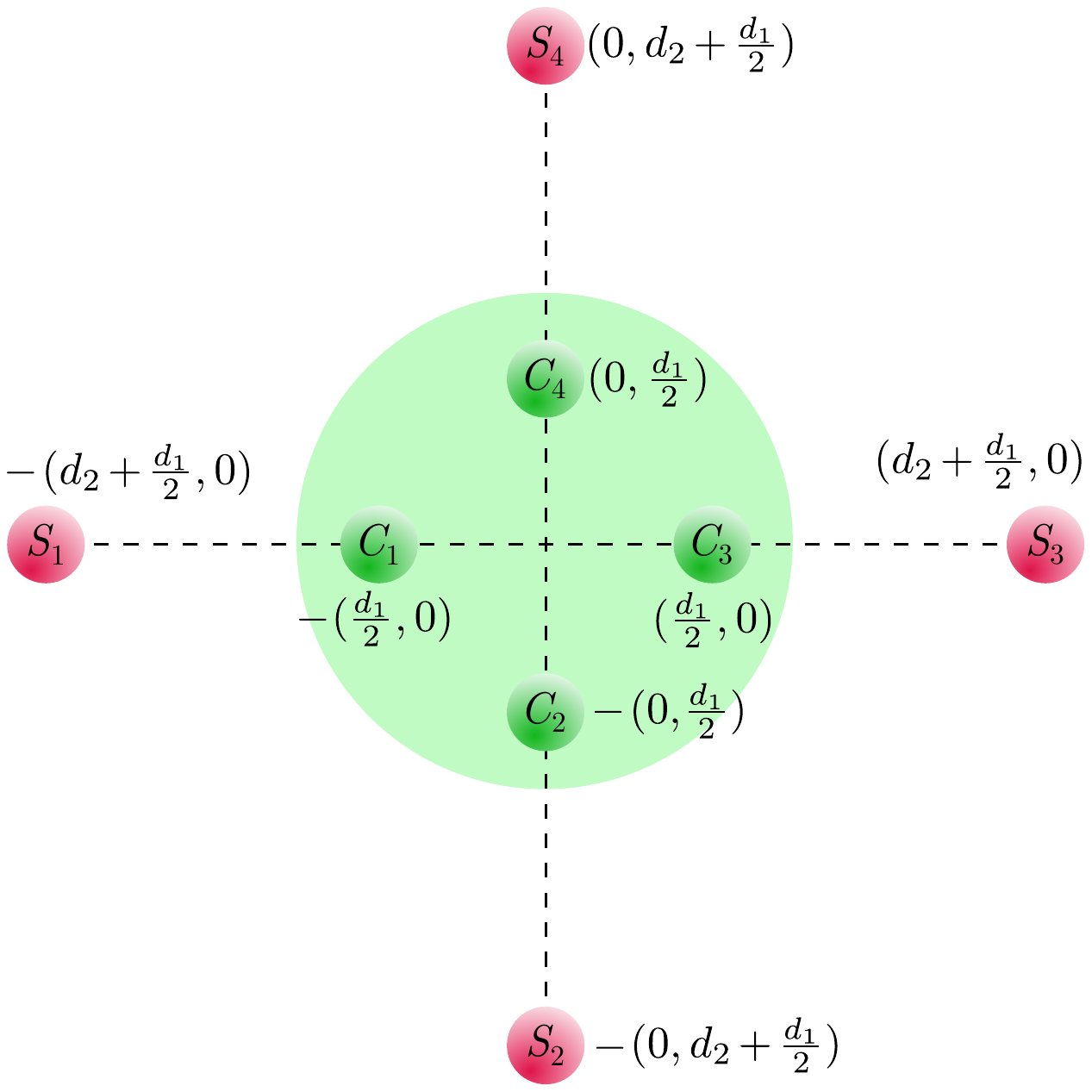} \label{fig:SM:2D}} \hspace{0.5cm}
\subfloat[]{\includegraphics[width=0.62\columnwidth]{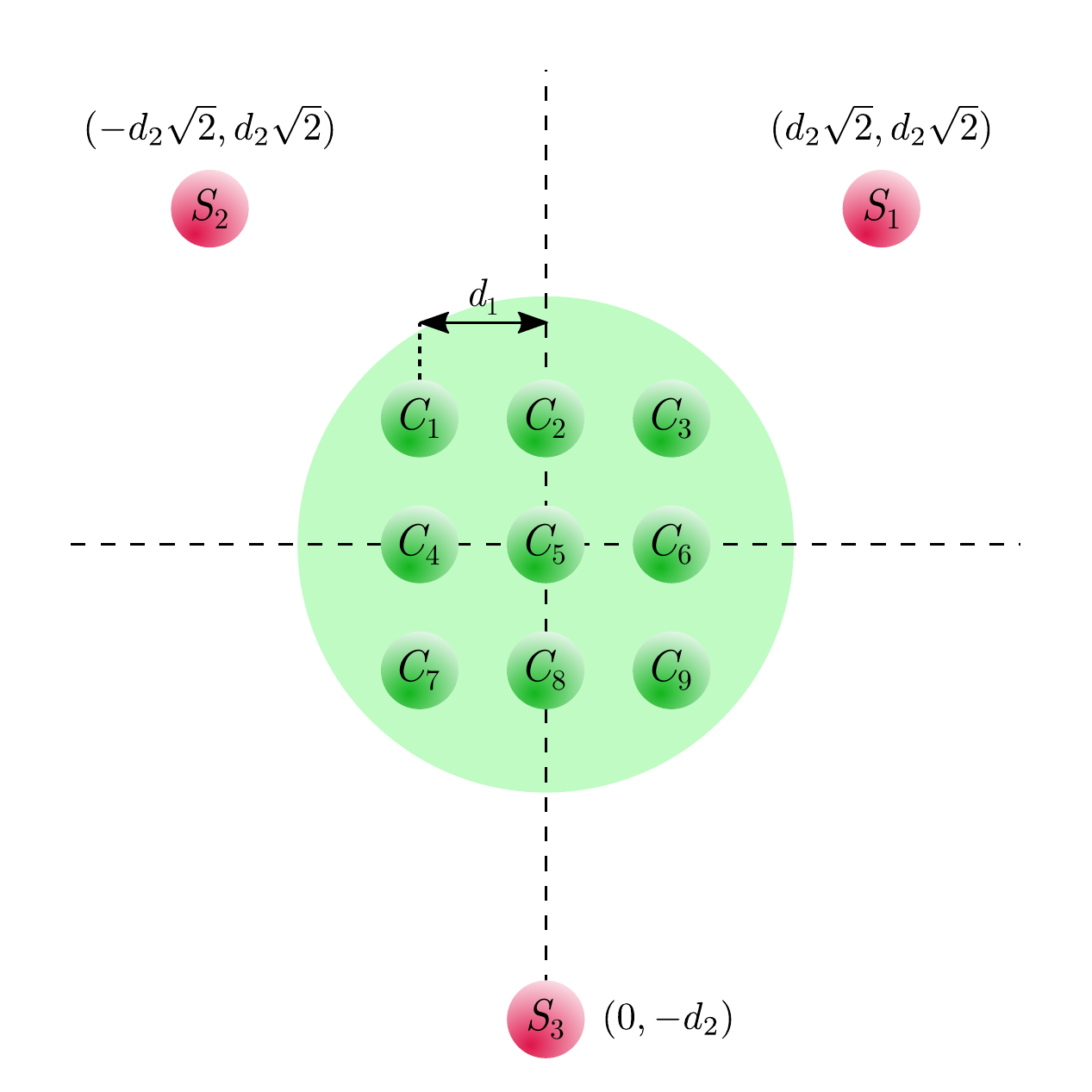} \label{fig:SM:2D:c}}
\caption{Schematic representation of the different distributions of the ions. Central circles (green) represent the qubits of the control system $C$, meanwhile, the circles of the extremes (red) correspond to the qubits of the target system $S$.}
\label{fig:SM}
\end{figure*}

Several experiments have been performed using systems of trapped ions with effective long-ranged, distant-dependent Ising interactions that are induced by spin-dependent optical dipole forces implemented by an applied laser field \cite{porras2004effective,pagano2020,joshi2020quantum}, where $f_{ij} = J | \boldsymbol{r}_{i} - \boldsymbol{r}_{j} |^{-\alpha}$ and $\alpha \approx 1$. This setup can also be used to demonstrate our approach, which we illustrate with two examples.

The minimal setting is a linear string of four ions, where $C$ is given by the two ions in the middle, while $S$ consists of the two ions at the ends of the chain, see Fig.~\ref{fig:SM:1D}. The internal state of $C$ allows one to couple selectively to either (or both) of the two other ions, and also to generate entanglement among the two outer qubits. Considering $\alpha = 1$ and a constant distance $d$ between qubits, we can couple $C$ to only one of the qubits by setting the control system into the logical sub-space given by $\boldsymbol{c} = (1, \, -0.5)^T$ with an interaction coupling of $\lambda_1 = 2.25 f^S_{12}$, $\lambda_2 = 0$, or $C$ can couple simultaneously to both qubits with the logical sub-space given by $\boldsymbol{c} = (1, \, 1)^T$ with an interaction coupling of $\lambda_{1, \, 2} = 4.5 f^S_{12}$, where $f^S_{12} = J / ( 3d )$ is the coupling strength of the inherent ion-ion interaction between the qubits of $S$. If there is a null interaction between qubit $s_1$ and qubit $S_2$, the time required to prepare a Bell state in the target system with the methods introduced in Sec.~\ref{Sec.Entanglement.generation.by.gate.sequence} and Sec.~\ref{Sec.Entanglement.generation.by.projective.measurements} are given in Equations~\eqref{eq:time.gates} and \eqref{eq:time.measurements}, where here correspond to
\begin{equation*}
\begin{aligned}
    t_g & = \frac{\pi d}{3 J} + \frac{\pi }{4 \omega} + \frac{5 \pi d}{4 J } \\
    t_m & = \frac{\pi d}{6 J} + \frac{\pi d}{4 J} + t_{\mathcal{M}}.
\end{aligned}
\end{equation*}

This example can easily be scaled up to chains of size $4n$, with $n$ ions at the borders forming $S$, and the remaining $2n$ ions in the middle forming $C$. See Appendix~\ref{app:1d:ions} for details.

The second example is a 2D cross setting as illustrated in Fig.~\ref{fig:SM:2D}, where a control system of four ions serves as an entanglement switch between the four outer qubits. Considering $\alpha = 1$ and a constant distance $d$ between qubits in each axis, we can simultaneously couple $C$ to selected spins with a coupling strength $\lambda_{1, \, 2} \approx 1.8 f^S_{12} $ for adjacent and $\lambda_{1, \, 3} =1.3 f^S_{13}$ for diametrically opposed pairs of qubits. Equally coupling $C$ to all the qubits the coupling strength is given by $\lambda_{1,2,3,4} \approx 6.5 \, \langle f^S_{ij} \rangle$. If there is a null interaction between qubit $S_1$ and qubit $S_2$, the time required to prepare a Bell state in the target system with the methods introduced in Sec.~\ref{Sec.Entanglement.generation.by.gate.sequence} and Sec.~\ref{Sec.Entanglement.generation.by.projective.measurements} are given in Equations~\eqref{eq:time.gates} and \eqref{eq:time.measurements}, where here correspond to
\begin{equation*}
\begin{aligned}
    t_g & = \frac{2 \pi d}{3.6 J \sqrt{2}} + \frac{ 15\sqrt{2} \pi d}{2 J} + \frac{\pi}{4 \omega} \\
    t_m & = \frac{\pi d}{3.6 J \sqrt{2}} + \frac{ 3\sqrt{2} \pi d}{2 J} + t_{\mathcal{M}}.
\end{aligned}
\end{equation*}
See Appendix~\ref{app:2d:ions} for details.

Note the effective coupling strength between spins can be increased using a larger control system that mediates interactions. In turn, suppressing some interactions and selectively coupling to specific spins reduces the coupling strength.

\section{Position noise in \textit{S}}
\label{Sec.Position.Noise}

An important feature of our setting is that it still can be used in the presence of noise in the position of the qubits, at the price of a slightly reduced fidelity of the prepared states. More importantly, the effects of position noise can be reduced by increasing the size of $C$. Moreover, with a larger control system, we can implement techniques introduced in \cite{hamann2021approximate}, which allow us to just assume that each qubit is located within a certain region instead of knowing its position exactly. This reduces the effective coupling, yet leads to higher fidelity of the prepared states.

\subsection{Noise model}
\label{Sec.Position.Noise.0}

First, we analyse how noise in the position of the target qubits affects our setting. We consider that the position of each qubit is given by a normal probability distribution, i.e., the probability of finding qubit $S_i$ in a two-dimensional region $R$ is given by
\begin{equation*}
    \int_R \rho_i \left( \boldsymbol{r} \right) \mathrm{d} \boldsymbol{r},
\end{equation*}
with
\begin{equation*}
    \rho_i \left( \boldsymbol{r} \right) = \frac{1}{2\pi \sigma^2} \, e^{ -\frac{ \left( \boldsymbol{r} - \bar{ \boldsymbol{r}}^S_i \right)^2 }{2 \sigma^2} },
\end{equation*}
where $\bar{\boldsymbol{r}}^S_i$ is the expectation value of its position, and $\sigma^2$ is the variance of the distribution.

As the position of the qubits is not completely determined, it is convenient to define the scalar field $\lambda(\boldsymbol{r})$ which corresponds to the interaction coupling between a virtual qubit at position $\boldsymbol{r}$ and $C$. It is given by
\begin{equation*}
    \lambda (\boldsymbol{r}) = \sum_{i = 1}^N c_i \, \frac{ J }{\left| \boldsymbol{r} - \boldsymbol{r}^C_i \right|}= \boldsymbol{f} (\boldsymbol{r}) \cdot \boldsymbol{c},
\end{equation*}
where
\begin{equation}
\label{eq:vectorfield}
    \boldsymbol{f}(\boldsymbol{r}) = \left( J \left| \boldsymbol{r} - \boldsymbol{r}^C_1 \right|^{-1}, \dots, J \left| \boldsymbol{r} - \boldsymbol{r}^C_N \right|^{-1} \right)
\end{equation}
is a vector field where component $f_i(\boldsymbol{r})$ corresponds to the physical coupling strength between the virtual qubit at $\boldsymbol{r}$ and qubit $C_i$.

Given a certain target interaction pattern $\boldsymbol{\lambda}^t = (\lambda^t_1, \dots, \lambda^t_n)^T$, we still can obtain the corresponding logical subspace by assuming that each qubit of $S$ is located at its mean value and computing the interaction matrix, i.e., $ ( \boldsymbol{F} )_{ij} = J | \boldsymbol{r}^C_i - \bar{ \boldsymbol{r} }^S_j|^{-1}$ and $\boldsymbol{c} = \boldsymbol{F}^{-1} \cdot \boldsymbol{\lambda}^t$. However, once a logical subspace is established, the actual interaction pattern depends on the exact location of the qubits, i.e., $\boldsymbol{\lambda}(\{\boldsymbol{r}^S_i\})$, whereby construction it fulfils that $\boldsymbol{\lambda}(\{ \bar{\boldsymbol{r}}^S_i \} ) = \boldsymbol{\lambda}^t$, and hence, $\boldsymbol{\lambda}$ is given by a probability distribution what yields a noisy evolution of the system. The probability distribution of $\lambda_i$ is given by $\rho(\lambda_i) = \rho_i (\boldsymbol{r}) \, (\mathrm{d}\boldsymbol{r}/\mathrm{d} \lambda_i)$, and hence, an initial state $\varrho(0)$ evolves as
\begin{widetext}
\begin{equation}
\label{eq:noisy:evolution}
    \varrho (t) = \int U \! \left( \{\boldsymbol{r}_k \}, t \right) \, \varrho(0) \; U^{\dagger} \! \left( \{\boldsymbol{r}_k \}, t \right) \, \rho_1 (\boldsymbol{r}_1) \cdots \rho_n (\boldsymbol{r}_n) \, \mathrm{d} \boldsymbol{r}_1 \cdots \mathrm{d} \boldsymbol{r}_n,
\end{equation}
\end{widetext}
where
\begin{equation}
\label{eq:Ur}
    U \left( \{\boldsymbol{r}_k\}, t \right) = e^{-\ti \sum_{k = 1}^n \lambda (\boldsymbol{r}_k) \, Z^C Z^S_k \, t }.
\end{equation}
\subsubsection{2D set-up: 8 ions with position fluctuation}

To see how position noise affects the performance of our setting, we consider this noisy model in the particular example of Fig.~\ref{fig:SM:2D} without qubit $S_4$, i.e., $S = \{S_1, S_2, S_3\}$. For that, we analyse the preparation of the maximally entangled state
\begin{equation}
\label{eq:targetstate}
    \ket{\psi} = \frac{1}{2}\big( \ket{00} + e^{\ti \pi/2 } \ket{01} + e^{\ti \pi/2 } \ket{10} + \ket{11} \big)
\end{equation}
with qubits $S_1$ and $S_2$, and we calculate the fidelity of resulting state $\varrho_\psi$ with the target state $\left|\psi\right\rangle\!\left\langle\psi\right|$.

We consider the method detailed in Sec.~\ref{Sec.Entanglement.generation.by.gate.sequence}, which consist in applying the gates sequence $U \exp\{\ti \frac{\pi}{4} X^C \} U^{\dagger}$ with
\begin{equation*}
    U = e^{-\ti \, \frac{\pi}{4} Z^C \left( Z^S_1 + Z^S_2 \right) }
\end{equation*}
to the initial state $\ket{+}_C\ket{+++}_S$. Due to the noise, the evolution of the state is given by Equations \eqref{eq:noisy:evolution} and \eqref{eq:Ur}, with $n = 3$, $t = \pi/(4\lambda_{\max}^{(1,2)})$, and $\lambda (\boldsymbol{r}_k) = \boldsymbol{f}(\boldsymbol{r}_k) \cdot \boldsymbol{c}^{(1,2)}$. Notice, that qubit $S_3$ must be also taken into account as due to the noise in its position, it is not totally decoupled from $C$.

The resulting state is a mixed state $\varrho_{\psi}$ that correlates $C$ and $S$. Therefore, the fidelity with the target state $\left|\psi\right\rangle\!\left\langle\psi\right|_{S_1 S_2}$ is given by
\begin{equation*}
    F = \langle \psi | \, \text{tr}_{C S_3} ( \varrho_{\psi} ) \, \ket{\psi}.
\end{equation*}
In Table~\ref{tab:1}, we show the fidelity of the resulting state for different values of the standard deviation $\sigma$ in the position of the qubits. Observe, that if the qubit is trapped with a small error the setting still allows one to prepare highly entangled states. On the other hand, when the error in the trap increases the fidelity of the prepared state drops drastically. In that case, by increasing the number of qubits in $C$ we can reduce the effect of noise, as we show in the next section.

\renewcommand{\arraystretch}{1.4}
\begin{table}[h!]
\centering
\begin{tabular}{|cc|} \hline
$\quad \sigma \quad$ & $F$ \\ \hline
$\quad 0.15 d \quad$ & $\quad 0.918986 \quad$ \\
$\quad 0.10 d \quad$ & $0.965297$ \\
$\quad 0.05 d \quad$ & $0.991552$ \\
$\quad 0.01 d \quad$ & $0.999665$ \\ \hline
\end{tabular}
\caption{\label{tab:1}Fidelity of the state of qubits $S_1$ and $S_2$, resulting from the preparation of the Bell state given in Eq. \eqref{eq:targetstate} with the method of Sec.~\ref{Sec.Entanglement.generation.by.gate.sequence}, for different values of the standard deviation $\sigma$ of the position of the qubits of $S$, where the spatial distribution of the qubits is shown in Fig.~\ref{fig:SM:2D} with $d=d_1=d_2$.}
\end{table}

\subsection{Reducing noise effects}

In this section, we show how we can protect our setting from the effect of noise on the positions of the target system by increasing the number of qubits in the control system. This is achieved by setting the effective couplings with each qubit for a certain region instead of only for the exact positions.

As we showed above, the interaction coupling between $C$ and a virtual qubit a position $\boldsymbol{r}$ is given by $\lambda(\boldsymbol{r}) = \boldsymbol{f}(\boldsymbol{r}) \cdot \boldsymbol{c}$, where $\boldsymbol{f}(\boldsymbol{r})$ is defined in Eq.~\eqref{eq:vectorfield}. So far, to find the logical subspace that generates a target interaction pattern $\boldsymbol{\lambda}^t$, we assume that each qubit is in a particular known position, i.e., we find the vector $\boldsymbol{c}$ that fulfils $\lambda_i^t = \boldsymbol{f}(\bar{\boldsymbol{r}}^S_i) \cdot \boldsymbol{c} \; \forall \, i$. However, when the position of each qubit is given by a probability distribution with a non-negligible variance, our assumption of its position can significantly decrease the fidelity of the resulting state, as we showed in the previous section. To reduce the impact of this kind of noise, we enhance our setting by considering that each qubit $S_i$ is located in a region $R_i$ around the expected value of its position $\bar{\boldsymbol{r}}^S_i$. Then, we establish a logical subspace for $C$ that generates the target interaction coupling $\lambda_i^t$ for all region $R_i$. We achieve that by implementing the two methods introduced in \cite{hamann2021approximate}.

\textbf{Method 1.} We call $R^{(n)}$ the region around a point $\boldsymbol{r}_0$ where the vector field $\boldsymbol{f}(\boldsymbol{r})$ can be approximated by its $n$-order Taylor expansion, i.e., assuming a two-dimensional space for the position of the qubits in $R^{(1)}$ we can approximate
\begin{widetext}
\begin{equation*}
    \boldsymbol{f} \left( \boldsymbol{r} \right) \, \approx \, \boldsymbol{f} (\boldsymbol{r}_0) + (x - x_0) \left( \frac{\partial \boldsymbol{f}}{\partial x} \right)_{\!\boldsymbol{r}_0} + (y - y_0) \left( \frac{\partial \boldsymbol{f}}{\partial y}\right)_{\!\boldsymbol{r}_0} ,
\end{equation*}
where $\boldsymbol{r}=(x,y)$, and hence, in $R^{(1)}$ the effective coupling with $C$ is given by
\begin{equation}
\label{app:eq:taylor}
    \lambda(\boldsymbol{r}) \, \approx \, \underset{(a) }{ \underbrace{ \boldsymbol{f} (\boldsymbol{r}_0) \cdot \boldsymbol{c} }} + (x - x_0) \underset{(b)}{ \underbrace{ \left[\left( \frac{\partial \boldsymbol{f}}{\partial x} \right)_{\!\boldsymbol{r}_0 } \cdot \boldsymbol{c} \, \right] }} + (y - y_0 )\underset{(c)}{ \underbrace{ \left[\left( \frac{\partial \boldsymbol{f}}{\partial y}\right)_{\!\boldsymbol{r}_0} \cdot \boldsymbol{c} \, \right] }}.
\end{equation}
\end{widetext}
Then, we can obtain an effective coupling $\lambda^t$ between $C$ and a virtual qubit in $R^{(1)}$, by finding a vector $\boldsymbol{c}$ that fulfills $(a) = \lambda^t$, and terms $(b) = (c) = 0$ in Eq.~\eqref{app:eq:taylor}. Therefore, when considering multiple target qubits with a noisy position, we can assume that each qubit is within its corresponding $R^{(1)}_i$ instead of in a particular position and then the logical vector $\boldsymbol{c}$ is such that fulfils $\widetilde{\boldsymbol{F}} \cdot \boldsymbol{c} = \boldsymbol{\lambda}^t$, i.e.,
\begin{equation*}
    \begin{pmatrix}
    \boldsymbol{f}(\bar{\boldsymbol{r}}^S_1) \\
    \vdots \\
    \boldsymbol{f}(\bar{\boldsymbol{r}}^S_n) \\
    (\partial_x \boldsymbol{f})(\bar{\boldsymbol{r}}^S_1) \\
    \vdots \\
    (\partial_y \boldsymbol{f})(\bar{\boldsymbol{r}}^S_n)
    \end{pmatrix}
    \begin{pmatrix}
    c_1 \\ c_2 \\ c_3 \\ \vdots \\ c_{N-1} \\ c_N
    \end{pmatrix}
    =
    \begin{pmatrix}
    \lambda_1^t \\ \vdots \\ \lambda_n^t \\ 0 \\ \vdots \\ 0
    \end{pmatrix}
\end{equation*}
where $dim( \widetilde{F} ) = 3n \times n$ (where $n$ is the number of qubits in $S$). Notice, that if $dim(\boldsymbol{c}) = 3n$ the system of equations can always be solved. This means if $S$ consist of $n$ qubits, $C$ has to contain $N = 3n$ qubits. If considering three-dimensional settings $\widetilde{F}$ has $4n$ rows and the required number of qubits in $C$ has to be given by $N = 4n$, as it has to be insensitive to the three spatial derivatives of each qubit location.

This provides us with a way of establishing an effective coupling with a target qubit that is within $R^{(1)}$. If the position noise is small enough to fulfil this assumption, the effect of the noise is significantly reduced. However, if the position noise is too large it may not be enough to assume that $S_i$ is within $R_i^{(1)}$. In this case, we assume that the qubits are located in a larger region $R^{(n)}$ and we establish the coupling for all that region. For that, field $\boldsymbol{f}(\boldsymbol{r})$ has to be expanded in its Taylor series until order $n$, and in the same way, as we did for $R^{(1)}$, we have to find a vector $\boldsymbol{c}$ that eliminates all the contributions except the zero order, i.e.,
\begin{equation*}
\begin{aligned}
    \boldsymbol{f}(\boldsymbol{r}_0) \cdot \boldsymbol{c} & = \lambda \\
    \left( \frac{\partial^k\boldsymbol{f}}{(\partial x)^{r} ( \partial y )^{k-r}} \right)_{\boldsymbol{r}_0} \cdot \boldsymbol{c} & = 0
\end{aligned}
\end{equation*}
for $0 \leq r \leq k$ and $1 \leq k \leq n$. In this way, if one can assume that the target qubits are located in finite regions that do not overlap, the effect of noise can be counteracted by considering a large enough control system. In particular, considering a two-dimensional position space the number of constraints that vector $\boldsymbol{c}$ has to fulfil increase quadratically with $n$, and therefore, the size of $C$.
\begin{figure}
    \centering
    \subfloat[]{\includegraphics[width=0.25\columnwidth]{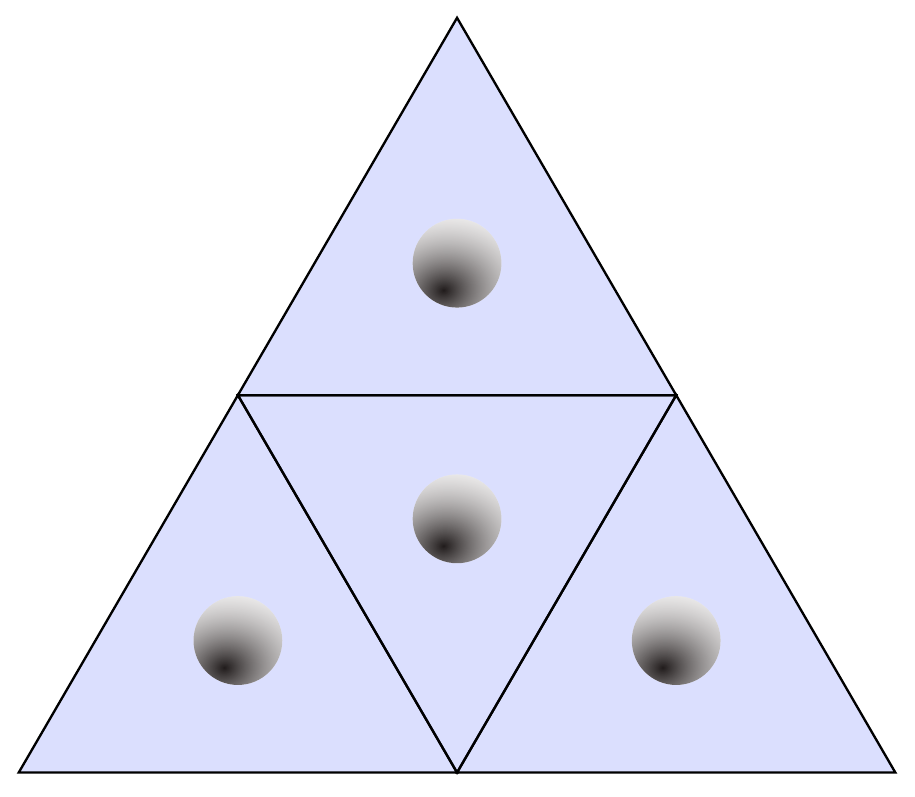} \label{fig:discret:a}}
    \subfloat[]{\includegraphics[width=0.25\columnwidth]{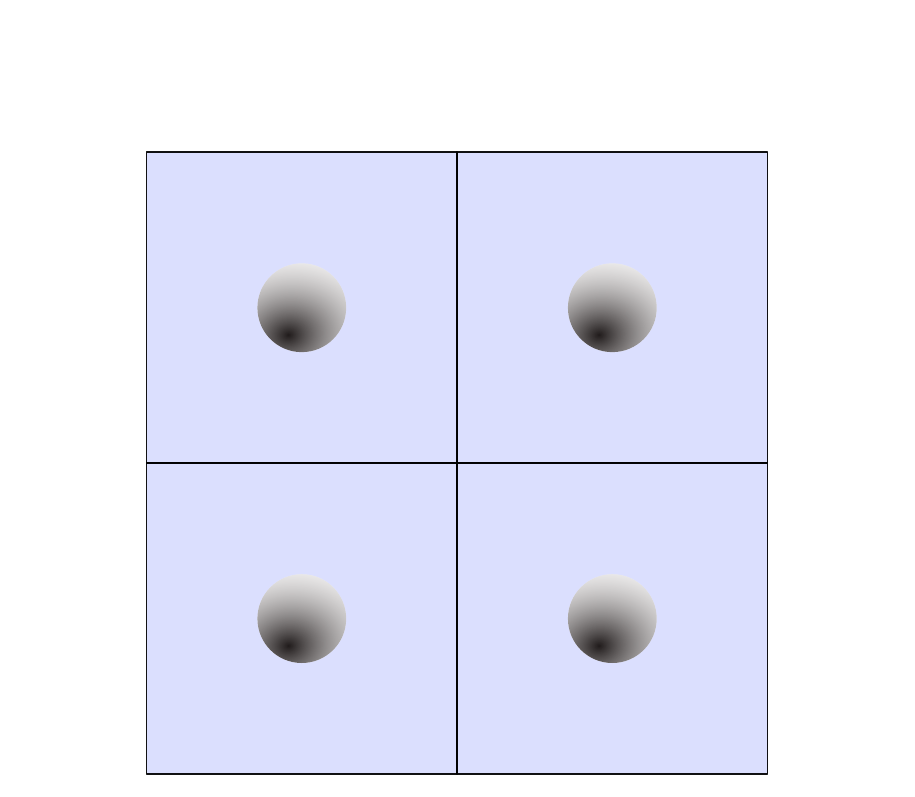}}
    \subfloat[]{\includegraphics[width=0.25\columnwidth]{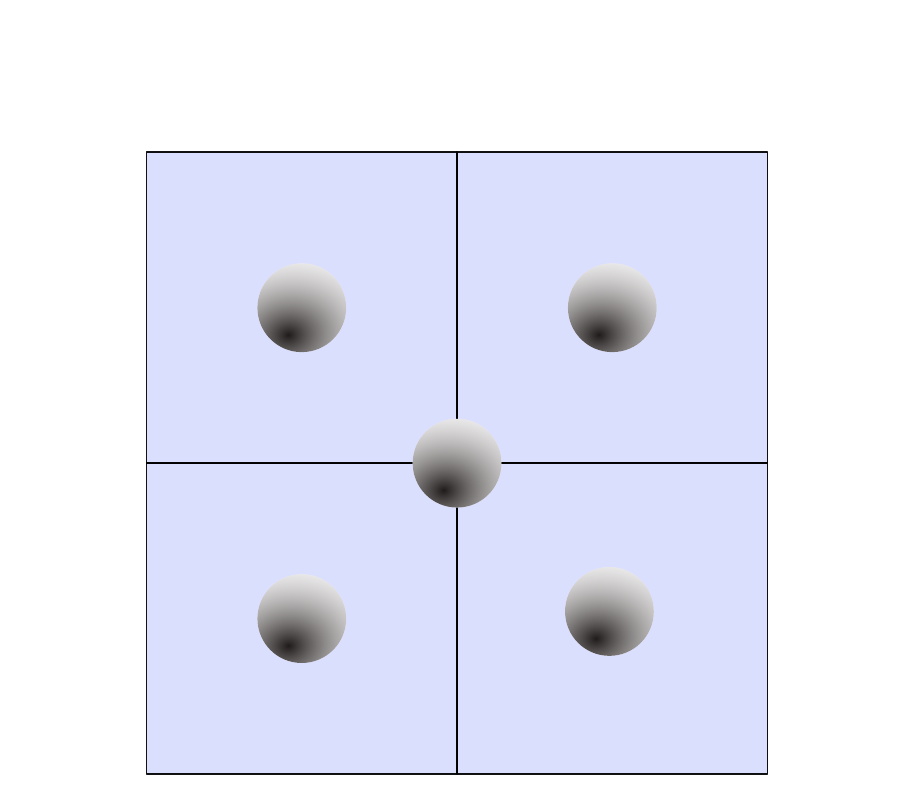}}
    \subfloat[]{\includegraphics[width=0.25\columnwidth]{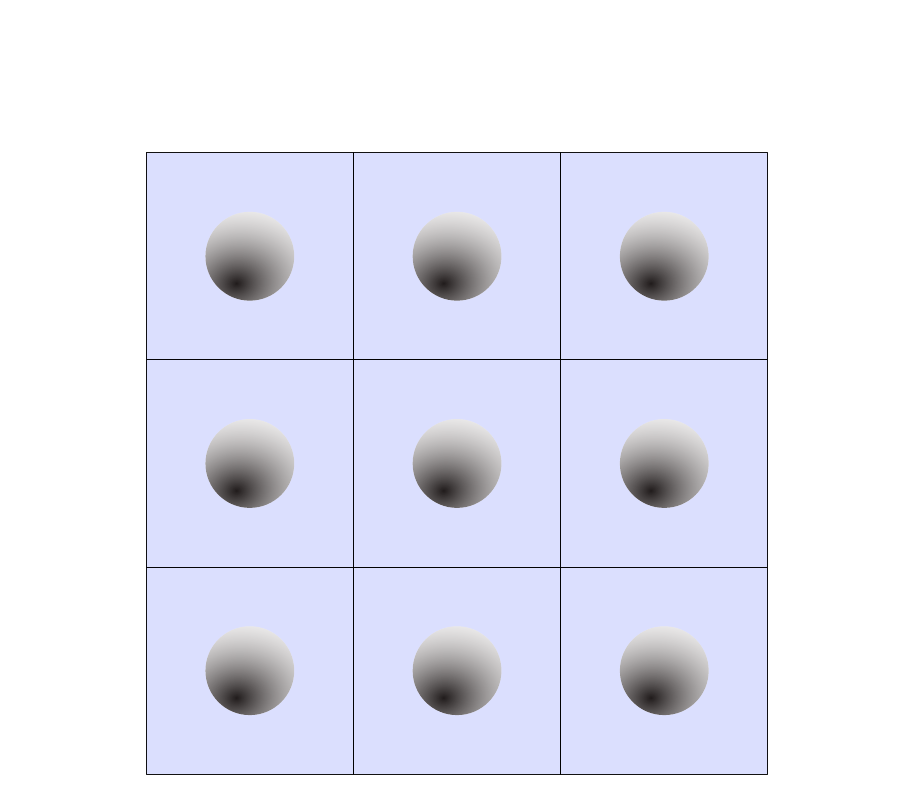}}
    \caption{\label{fig:discret} Four different ways of discretise region $R$. Black circles represent virtual qubits.}
\end{figure}

\textbf{Method 2}. We also can establish an approximated effective coupling for region $R$ by discretising it. Given a region $R$ we divided it in $k$ subregions $R^{(i)}$, and we consider a virtual qubit $v_i$ in the center of each subregion $\boldsymbol{r}_{v_i}$, see Fig.~\ref{fig:discret}. To find a logical subspace that couples with a target qubit within region $R$ with a certain effective coupling $\lambda$, we just have to obtain a subspace that yields the same coupling with all the virtual qubits, i.e.,
\begin{equation*}
    \begin{pmatrix}
    c_1 \\ \vdots \\ c_N
    \end{pmatrix}
    =
    \begin{pmatrix}
    \boldsymbol{f}(\boldsymbol{r}_{v_1}) \\ \vdots \\ \boldsymbol{f}(\boldsymbol{r}_{v_k})
    \end{pmatrix}^{-1}
    \begin{pmatrix}
    \lambda \\ \vdots \\ \lambda
    \end{pmatrix}
\end{equation*}
Note that with this method, to couple $C$ with $n$ target qubits, the control system has to contain $kn$ qubits where $k$ is the number of virtual qubits in each region.

This approach guarantees a constant coupling strength whenever the physical qubit is at the position of one of the virtual qubits. By continuity, the coupling strength is also approximately the same in the neighbourhood of the positions, and hence in the whole region.

\subsubsection{Two-dimensional example of noise reduction}

\renewcommand{\arraystretch}{1.4}
\begin{table*}
\centering
\begin{tabular}{|ccccc|} \hline
$\sigma$ & $F_0$ & $F_1$ & $F_2$ & $F_3$ \\ \hline
$\quad 0.25 d \quad$ & $0.934582$ & $0.948641$ & $0.981530$ & $0.990489$ \\
$0.20 d$ & $\quad 0.959314 \quad$ & $\quad 0.967918 \quad$ & $0.993573$ & $\quad 0.996809 \quad$ \\
$0.15 d$ & $0.977671$ & $0.982306$ & $\quad 0.998282 \quad$ & $0.999155$ \\
$0.10 d$ & $0.990255$ & $0.992248$ & $0.999701$ & $0.999853$ \\ \hline
\end{tabular}
\caption{\label{tab:2}Fidelity of the prepare state of with qubits $S_1$ and $S_2$, with the target state given in Eq.~\eqref{eq:targetstate} with the method of Sec.~\ref{Sec.Entanglement.generation.by.gate.sequence}, for different values of the standard deviation $\sigma$ of the position of the qubits of $S$, where the spatial distribution of the qubits is shown in Fig.~\ref{fig:SM:2D:c} with $d=d_1=d_2$. $F_i$ consider the state obtained by using $\boldsymbol{c}_i$.}
\end{table*}

\begin{table*}
\centering
\begin{tabular}{|ccc|} \hline
$\sigma$ & $\langle \lambda_{1,2} \rangle$ & $\boldsymbol{c}_3$ \\ \hline
$\;\; 0.25 d \;\;$ & $\;\;0.0139\;\;$ & $\;\; (-0.0970, 0.3703, -0.0966, 0.3827, -1, 0.3814, -0.0331, 0.1508, -0.0324) \;\;$ \\
$0.20 d$ & 0.0136 & $(-0.0945, 0.3675, -0.0941, 0.3788, -1, 0.3775, -0.0307, 0.1504, -0.0301)$ \\
$0.15 d$ & 0.0133 & $(-0.0927, 0.3654, -0.0923, 0.3756, -1, 0.3748, -0.0292, 0.1506, -0.0287)$ \\
$0.10 d$ & 0.0131 & $(-0.0913, 0.3639, -0.0911, 0.3739, -1, 0.3735, -0.0285, 0.1513, -0.0282)$ \\ \hline
\end{tabular}
\caption{\label{tab:3}Logical vector used to couple to $S_1$ and $S_2$ with method 2, where the spatial distribution of the qubits is shown in Fig.~\ref{fig:SM:2D:c} with $d=d_1=d_2$.}
\end{table*}

We analyse the effect of position noise in a particular example. We compare the fidelity of the prepared state in the minimal size setting with different options to reduce the noise effects. In total, we consider three models. Model 0 consists of the minimal-size control system, which is the one where $C$ contains the same number of qubits as $S$. Model 1 is given by a setting where the number of qubits in $C$ is three times larger than in $S$. Models 2 and 3 are given by the same setting used in model 1 but by generating the effective coupling of $C$ around a certain region $R_i^{(1)}$ for each qubit by using method 1 and method 2 respectively. We denote as $F_k$ the prepared state fidelity and $\boldsymbol{c}_k$ the logical subspace of $C$ of model $k$.

In particular, we consider a target system $S$ of 3 qubits where the qubits are distributed as shown in Fig.~\ref{fig:SM:2D:c} where we take $d_2 = 3 d_1$. The qubits of $C$ are also spatially distributed as shown in Fig.~\ref{fig:SM:2D:c}, but for the minimal-size setting, we only consider qubits $\{C_1, C_3, C_8\}$. We then assume the preparation of the Bell state given in Eq.~\eqref{eq:targetstate} between $S_1$ and $S_2$ with the method detailed in Sec.~\ref{Sec.Entanglement.generation.by.gate.sequence}.

For that, we have to generate the target interaction pattern given by $\boldsymbol{\lambda}^t = (1,\,1,\,0)^T$. We compute the logical subspace as $\boldsymbol{c} = \boldsymbol{F}^{-1} \cdot \boldsymbol{\lambda}^t$ for the mean value in the position of the qubits. We obtain: for the minimal-size setting, i.e., model 0,
\begin{equation*}
    \boldsymbol{c}_0^{(1,2)} = \Big( 1, \, 1, \, -4/\sqrt{17} \Big)^T
\end{equation*}
with $\langle \lambda_{1,2} \rangle = 0.675 J/d_1$. For model 1
\begin{equation*}
\begin{aligned}
    \boldsymbol{c}_1^{(1,2)} = \Big( 1, \, 1, \, 1, \, 1, \, 0.0807, \, 1, \, -1, \, -1, \, -1 \Big)^T
\end{aligned}
\end{equation*}
with $\langle \lambda_{1,2} \rangle = 1.265 J/d_1$. For model 2 we obtain the logical subspace as $\boldsymbol{\boldsymbol{c}} = \widetilde{\boldsymbol{F}}^{-1} \cdot \tilde{\boldsymbol{\lambda}}^t$, where $\tilde{\boldsymbol{\lambda}}^t = (\lambda^t_1,\lambda^t_2,\lambda^t_3,0,\dots,0)$, and we obtain
\begin{equation*}
\begin{aligned}
    \boldsymbol{c}_2^{(1,2)} = \Big( -0.4842, \, 0.5667, \, -0.4842, \, 1, & \\
    -0.1852, \, 1, \, -1, \, 0.8339, & \, -1 \Big)^T
\end{aligned}
\end{equation*}
with $\langle \lambda_{1,2} \rangle = 0.098 J/d_1$. For model 3 we assume a virtual qubit in the expected value of the position, and the other two around the first one as shown in Fig.~\ref{fig:discret:a} at a distance of $\sigma$. therefore, the logical subspaces used depend on $\sigma$ and are shown in Table~\ref{tab:3}.

We compute the fidelity $F$ of the prepared state by using the four different logical subspaces, $\boldsymbol{c}_0$, $\boldsymbol{c}_1$, $\boldsymbol{c}_2$ and $\boldsymbol{c}_3$, see Table~\ref{tab:2}. Observe the lowest fidelity is obtained for the minimal-size control system, i.e., if $C$ only contains three qubits ($C_1$, $C_2$ and $C_8$). Then, if $C$ contains nine qubits, a stronger coupling can be obtained and the effect of noise is reduced, i.e., the fidelity of the output state is larger. Finally, considering nine qubits in $C$ and the coupling for each qubit is established in a region $R_i^{(1)}$, the effect of noise is strongly reduced even though a much weaker coupling is obtained. Therefore, in this last case, the evolution time required to produce the state is larger.

\section{Conclusion and outlook}
\label{Sec.Conclusion}

We have introduced a method to generate entanglement in systems without quantum control in a remotely controlled way. The main tool is the effective control of distance-dependent always-on interactions by the choice of the internal state of a high-dimensional control system. By using the control system as an effective two-level system, i.e., as a single logical qubit, we can selectively couple it to outside systems, and use built-up entanglement to mediate interactions, perform gates and prepare entangled states on a remote system. While the latter has been utilized in different contexts \cite{spee2013remote}, the control of effective couplings by choice of internal states only provides new and interesting possibilities for the design of quantum processors or entanglement switches. In Appendix~\ref{app:comparizing} we also compare our approach to more restricted techniques introduced in \cite{albertini2018controllability,albertini2020subspace,albertini2021subspace,d2021dynamical}.

We point out that the techniques to deal with spatially correlated noise processes outlined in \cite{sekatski2020optimal,W_lk_2020} in the context of distributed quantum metrology can be utilized here as well. That is, we can use our control system also in such a way that noise processes with spatial correlation (e.g., a constant but fluctuating global field) can be fully suppressed in the target system while maintaining the full functionality of our remote entanglement preparation scheme. This only requires a slightly enlarged control system. In addition, noise within the control system can be treated similarly. However, that is a much more complex problem that we treat in \cite{ferran:noisygates}.

Finally, we point out that we use our control system as an artificial two-level system, where we can control the interactions with the surroundings by choice of internal state. One can use a collection of such artificial two-level systems to realize a programmable quantum simulator \cite{rieradur:simulation} in the presence of some distance-dependent, always-on coupling.

\begin{acknowledgments}
This work was supported by the Austrian Science Fund (FWF) through projects No. P30937-N27, No. P36009-N and No. P36010-N. Finanziert von der Europ\"aischen Union - NextGenerationEU.
\end{acknowledgments}

\bibliographystyle{apsrev4-1}
\bibliography{QuantumSwitch.bib}

\onecolumn
\appendix

\section{Comparison with previous results on remote control}
\label{app:comparizing}

Other control device techniques have been developed in recent years. Here, we point out the main differences with our approach, including the considered problems, methods, and possible applications.

In our approach, we consider the controllability of an ensemble of spins utilizing an auxiliary quantum system in the presence of an interaction with some (arbitrary) distance dependence. Indeed, this is similar to the scenario considered in [3-6], where criteria to add control to a remote quantum system were studied. However, the approaches are fundamentally different as we now explain.

In Refs. [3-6] symmetry assumptions are crucial. In these works, the dynamical Lie algebra of the setting is analyzed to obtain controllable subspaces, by studying the symmetries in the target system. On the other hand, our setting uses the distance dependence of the Ising interaction between the qubits to selectively choose the coupling of our control system. There is no need for any underlying symmetry, and in fact, the method is also applicable for other interactions, in particular also with other distance dependencies - or even other types of interactions if one allows for dynamical control as we point out below. As long as interactions show some kind of distance dependence, we provide methods to remotely generate entanglement in a target system without control there. In this case, we can perform any computational diagonal unitary in our target system and therefore generate any state of the form

\begin{equation*}
    \ket{\psi} = \sum_{i_1,\dots,i_n=0}^1 e^{\ti \, \theta_{i_1\cdots i_n}} \ket{ i_1, \dots, i_n},
\end{equation*}
what includes any stabilizer or graph state.

To summarize, we point out the advantages of our method over previous approaches, and novel contributions:

\begin{itemize}
\item We use an auxiliary control system that consists of multiple qubits. Local control of individual single qubits is sufficient to obtain not only full control within the control system (using the always-on self-interaction) but also to achieve remote control in an outside system that is not accessible.

\item We explicitly show how to remotely generate multipartite entanglement efficiently. We explicitly provide the states and control sequences that are required for particular settings.

\item Our methods are applicable in asymmetric situations and solely rely on the distance dependence in the underlying interaction that can be arbitrary.

\item Our approach can be extended to XY and XYZ interactions using (fast) local control in the control system only, as we now point out in the next section.

\item We provide a way to deal with noise and imperfections. In particular, we show how one can deal with thermal position fluctuation in the target systems, and still obtain entangled target states with high fidelity. Increasing the size of the control system allows one to increase fidelity.
\end{itemize}

\section{Heisenberg interaction model}
\label{app:heisemberg interaction}

In this appendix we show how our setting still can be implemented if a general XYZ-type interaction (including e.g. Heisenberg interaction) is given between the qubits, i.e., the interaction between $C$ and $S$ is given by
\begin{equation*}    
    H^{CS} = \sum_{\substack{1 \leq i \leq N \\ 1 \leq j \leq n}} \alpha^{CS}_{i j} \, X^C_i X^S_j + \beta^{CS}_{i j} \, Y^C_i Y^S_j + \gamma^{CS}_{i j} \, Z^C_i Z^S_j .
\end{equation*}
The analysis can not be straightforwardly extended to deal with such kinds of interactions directly. However, local operations in the control system suffice to modify the interactions between $S$ and $C$ to the ZZ-coupling we consider in the main text. This makes our method also applicable for the more general interactions and hence to a much larger class of systems and set-ups.

The basic idea is to use fast local control to eliminate some of the interaction of the Hamiltonian. This can be done by fast intermediate control pulses in $C$ only, which leads to a negative sign for $XX$ and $YY$ terms, and allows one to eliminate them by altering between the modified and unperturbed evolution. In other words, we define
\begin{equation*}
    U(t) = e^{-\ti H^{CS} t}
\end{equation*}
and
\begin{equation*}
    \tilde{U}(t) = Z^C_1 \cdots Z^C_N \, e^{-\ti H^{CS} t} \, Z^C_1 \cdots Z^C_N,
\end{equation*}
and by alternating them we reproduce the pairwise ZZ interaction, i.e.,
\begin{equation*}
    \lim_{k \to \infty} \left[ \tilde{U} \left( t / 2k \right) U \left( t / 2k \right) \right]^k = e^{-\ti H_z^{CS} t}
\end{equation*}
where
\begin{equation*}
    H_{z}^{CS} = \sum_{\substack{1 \leq i \leq N \\ 1 \leq j \leq n}} \gamma^{CS}_{i j} Z^C_i Z^S_j .
\end{equation*}

\section{Entanglement generation by generating multi-qubit \textit{Z} gates}
\label{app:multi:Z}

Multi-qubit gates with arbitrary phases can be mediated in any subset $S'\subseteq S$ by manipulating $C$. Applying first a maximally entangling operation between the qubits and the control system, given by $U^{S'}$, see Eq.~\eqref{eq:US}, followed by an X-rotation of the logical qubit and again by the same entangling operation, one produces a multi-qubit interaction between the logical qubit and the qubits in $S'$, i.e.,
\begin{equation*}
    U^{S'} \, e^{-\ti \, \omega X^C } \, U^{\dagger} = e^{-\ti \, \omega G^C Z^S_1 \! \cdots Z^S_N },
\end{equation*}
where if $C$ couples to the $n'$ qubits in $S'$, the unitary $G$ is given in Eq.~\eqref{eq:G}. Note that, $G \in \{ \pm X, \pm Y \}$, and leaving the state of $C$ in an eigenstate of $G^C$, one obtains an effective multiqubit Z-interaction in $S$ which can, e.g., be used to generate GHZ states (up to local unitary operations) directly by choosing $\omega = \pi / 4$.

Alternatively, one can use this method on all subsets $S'$ of qubits sequentially. There are $2^n - 1$ such subsets, and one can freely choose the interaction strength $\omega_i$ for each subgroup. Hence, with a proper choice of induced interaction phases $\{ \omega_i \}_{i=1}^{2^n-1}$, one can generate all states of the form Eq.~\eqref{eq:allstates}. Notice that the phases $\{ \omega_i\}$ can be determined from $\{\theta_{1, \dots, n}\}$ of Eq.~\eqref{eq:allstates} by solving a system of linear equations. These methods are however costly, as maximally entangling gates between $C$ and $S$ are required to produce even a small interaction, and at least four maximally entangling gates are used to generate a maximally entangled state in $S$.

\section{Entanglement generation by control rotation: details}
\label{app:gate:maentic:field}

Setting an interaction pattern that equally couples the control system with two qubits $S_i$ and $S_j$, and simultaneously applying a rotation to the logical qubit in $C$ in the X-direction, one can implement an effective ZZ-interaction between the qubits.

Such evolution is generated by
\begin{equation*}
    H_x = \lambda Z^C \big( Z^S_i + Z^S_j \big) + \omega X^C ,
\end{equation*}
and it transforms a general two qubits state $\ket{\psi} = \sum_{i,j=0}^1 \psi_{ij} \ket{ij}$ as
\begin{equation*}
\begin{aligned}
    e^{-\ti H_x t} \ket{+}_{C} \ket{\psi}_{S_i S_j} \\
    = \frac{1}{\sqrt{2}} \big( a_+ \psi_{00} \ket{0}_C \ket{00}_{S_i S_j} + b \, \psi_{01} \ket{0}_C \ket{01}_{S_i S_j} + b \, \psi_{10} \ket{0}_C \ket{10}_{S_i S_j} + a_- \psi_{11} \ket{0}_C \ket{11}_{S_i S_j} \\
    + \, a_- \psi_{00} \ket{1}_C \ket{00}_{S_i S_j} + \, b \, \psi_{01} \ket{1}_C \ket{01}_{S_i S_j} + b \, \psi_{10} \ket{1}_C \ket{10}_{S_i S_j} + a_+ \psi_{11} \ket{1}_C \ket{11}_{S_i S_j} & \big),
\end{aligned}
\end{equation*}
where
\begin{equation*}
\begin{gathered}
    a_{\pm} = \cos \left( \sqrt{\omega^2 + 4 \lambda^2} t \right) - \ti \, \frac{(\omega \pm 2 \lambda) \sin \left( \sqrt{\omega^2+ 4 \lambda^2} t \right)}{\sqrt{\omega^2 + 4 \lambda^2} } \\
    b = e^{-\ti \omega t}.
\end{gathered}
\end{equation*}
Note that $a_{\pm} = (-1)^k$ for $t = k \pi / \sqrt{ \omega^2 + 4 \lambda^2 }$ ($k \in \mathbb{N}$), and hence the control system factors-out leaving the state of $S_i$ and $S_j$ in an entangled state, i.e.,
\begin{equation*}
    e^{-\ti H_x \tau} \ket{+}_C \ket{\psi}_{S_i S_j} = \ket{+}_C \left( \psi_{00} \ket{00}_{S_i S_j} + e^{\ti 2 \phi } \psi_{01} \ket{01}_{S_i S_j} + e^{\ti 2 \phi } \psi_{10} \ket{10}_{S_i S_j} + \psi_{11} \ket{11}_{S_i S_j} \right)
\end{equation*}
where $\phi = \frac{\pi k}{2} ( 1 - \omega / \sqrt{\omega^2 + 4 \lambda^2} ) $. Since the final state of $S_i$ and $S_j$ can be written as $e^{-\ti \, \phi Z Z} \ket{\psi}$ an arbitrary effective ZZ-interaction is generated between $S_i$ and $S_j$.

For an initial state $\ket{\psi} = \ket{+\,+}$, setting $\lambda = \omega \sqrt{3} / 2$ and leaving the system to evolve a time $\tau = \pi / (2 \omega)$ the induced phase between the qubits is given by $\phi = \pi/ 4$ and therefore the qubits end up in a maximally entangled state, i.e.,
\begin{equation*}
    e^{-\ti \frac{\pi}{4} ZZ} \ket{++} = \frac{1}{\sqrt{2}} \big(\ket{00_y} + \ti \, \ket{11_y} \big)
\end{equation*}
where $\ket{0_y} = (\ket{0} + \ti \, \ket{1})/\sqrt{2}$ and $\ket{1_y} = (\ket{0} - \ti \, \ket{0})/\sqrt{2}$.

The same method can be used to prepare a locally unitary (LU) equivalent GHZ state with three qubits. The evolution generated by
\begin{equation*}
    H_x = \lambda Z^C \big( Z^S_i + Z^S_j + Z^S_k \big) + \omega X^C
\end{equation*}
leads to
\begin{equation*}
\begin{aligned}
    e^{-\ti H_x  t} \ket{+}_{C} \ket{+++}_{S_i S_j S_k} & = \frac{1}{4} \big( a_+ \ket{0}\ket{000} + b_+ \ket{0}\ket{001} + b_+ \ket{0}\ket{010} + b_- \ket{0} \ket{011} \\
    & + b_+ \ket{0}\ket{100} + b_- \ket{0}\ket{101} + b_- \ket{0}\ket{110} + a_- \ket{0}\ket{111} \\
    & + a_- \ket{1}\ket{000} + b_- \ket{1}\ket{001} + b_- \ket{1} \ket{010} + b_+ \ket{1}\ket{011} \\
    & + b_- \ket{1}\ket{100} + b_+ \ket{1}\ket{101} + b_+ \ket{1}\ket{110} + a_+ \ket{1}\ket{111} \big)
\end{aligned}
\end{equation*}
where
\begin{equation*}
\begin{aligned}
    a_{\pm} & = \cos \left( \sqrt{ \omega^2 + 9 \lambda^2} \, t \right) - \ti \, \frac{ \omega \pm 3 \lambda }{\sqrt{ \omega^2 + 9 \lambda^2}} \sin \left( \sqrt{ \omega^2 + 9 \lambda^2 } \, t\right) \\
    b_{\pm} & = \cos \left( \sqrt{ \omega^2 + \lambda^2} \, t \right) - \ti \, \frac{\omega \pm \lambda }{\sqrt{\omega^2 + \lambda^2 }} \sin \left( \sqrt{ \omega^2 + \lambda^2 } \, t\right).
\end{aligned}
\end{equation*}
Setting $\omega = \sqrt{5}\lambda/\sqrt{3}$ such that $\sqrt{ \omega^2 + 9 \lambda^2} =2 \sqrt{ \omega^2 + \lambda^2}$, and $\lambda = \sqrt{3} \pi / (2 \sqrt{2} t)$ such that $t \sqrt{ \omega^2 + \lambda^2}= \pi$ and $t \sqrt{ \omega^2 + 9 \lambda^2}= 2\pi$ one guarantees that the complex part of the coefficients cancels  and they become  $b_{\pm} = -1$ and $a_{\pm} = 1$. At thus time the control state factors out leaving the qubits in a LU equivalent graph state, i.e.,
\begin{equation*}
\begin{gathered}
    e^{-\ti \, \frac{\pi}{4} H_x \tau} \ket{+}_{C} \ket{+++}_{S_i S_j S_k} = e^{-\ti \, \frac{\pi}{4} \left( Z^S_i Z^S_j + Z^S_i Z^S_k + Z^S_j Z^S_k \right) } \ket{+}_C \ket{+++}_{S_i S_j S_k} \\
    = \frac{1}{2\sqrt{2}} \ket{+} \big( \ket{000} - \ket{001} - \ket{010} - \ket{011} - \ket{100} - \ket{101} - \ket{110} + \ket{111} \big).
\end{gathered}
\end{equation*}

\section{Entanglement generation by alternating interaction patterns}
\label{app:gate:sequence}

A simple way to entangle two qubits of the target system is to let $C$ and $S$ evolve sequentially under two different interaction patterns where each pattern couples $C$ to only one of the two qubits. With the appropriate locally (at $C$) altered patterns, a Bell state between the pair of qubits can be generated while the state of the control system factors out, e.g.,
\begin{equation*}
\begin{aligned}
    e^{-\ti \, \frac{\pi}{4} Z^C Z^S_i } e^{-\ti \, \frac{\pi}{4} Y^C Z^S_j } e^{-\ti \, \frac{\pi}{4} Z^C Z^S_i} \ket{+}_C \ket{+ +}_{S_i S_j} = \frac{1}{\sqrt{2}} \, \ket{-}_C \Big( \ket{0 0_y }_{S_i S_j} - \ti \, \ket{1 1_y }_{S_i S_j} & \Big) .
\end{aligned}
\end{equation*}
where $\ket{0_y} = (\ket{0} + \ti \, \ket{1})/\sqrt{2}$ and $\ket{1_y} = (\ket{0} - \ti \, \ket{1})/\sqrt{2}$. Notice that, the evolution generated by $\lambda Y^C Z^S_i $ can be obtained by applying a change of basis of the logical qubit,
\begin{equation*}
    e^{ \ti \, \frac{\pi}{4} X^C} \, e^{-\ti \lambda Z^C Z^S_i t } \, e^{-\ti \, \frac{\pi}{4} X^C} = e^{-\ti \lambda \, Y^C Z^S_i \, t }.
\end{equation*}
By repeating the protocol on different pairs of qubits in $S$, one can generate any graph state in up to some LU operations.

\section{Entanglement generation by Hamiltonian simulation}
\label{app:Hamiltonian:simulation}

Alternating the evolution generated by two Hamiltonians $H_1$ and $H_2$, one can approximate the evolution generated by the commutator $[H_1, \, H_2]$ for small $t$, as $e^{\ti H_1 t} e^{\ti H_2 t} e^{-\ti H_1 t} e^{-\ti H_2 t} \approx e^{- [H_1, H_2] \, t^2}$. One can use this technique to generate an effective three-qubit interaction in second order by manipulating the control system. Setting the two interaction patterns
\begin{equation*}
\begin{gathered}
    H_1 = \sum_{i = 1}^n \lambda_i X^C Z^S_i \\
    H_2 = \sum_{i = 1}^n \mu_i Y^C Z^S_i
\end{gathered}
\end{equation*}
one approximates the evolution given by the commutator
\begin{equation*}
    \left[ H_1, H_2 \right] = 2 \, \ti \sum_{i,j=1}^n \lambda_i \mu_j Z^C Z^S_i Z^S_j .
\end{equation*}
Then by leaving the state of $C$ in an eigenstate of $Z^C$, an effective ZZ pairwise interaction is generated in $S$ where the interaction coupling between qubits $S_i$ and $S_j$ is given by $2 ( \lambda_i \mu_j + \lambda_j \mu_i )$.

\section{1D set-up: 4 ions}
\label{app:1d:ions}

Assume four ions trapped in a 1D line where the separation between the ions is given by $d_1$, $d_2$ and $d_3$, as it is shown in Fig.~\ref{fig:SM:1D}, and consider the coupling between ions given by $J |\boldsymbol{r}_{i} - \boldsymbol{r}_{j} |^{-1}$. For this particular example, the inverse of the interaction matrix $\boldsymbol{F}$ is given by
\begin{equation*}
    \boldsymbol{F}^{-1} = \frac{1}{ J d_2 ( d_1 + d_2 + d_3 ) }
    \begin{pmatrix}
    d_1 (d_1 + d_2) (d_2 + d_3) & -d_1 d_3 (d_2 + d_3) \\
    - d_1 d_3 (d_1 + d_2) & d_3 (d_1 + d_2) (d_2 + d_3).
    \end{pmatrix}
\end{equation*}

In this scenario we have three possible interaction patterns:
\begin{enumerate}
    \item Coupling to qubit $S_1$: $$ \boldsymbol{\lambda}^{ (1) } = \lambda^{(1)} \big( 1, \, 0 \big)^T. $$
    \item Coupling to qubit $S_2$: $$\boldsymbol{\lambda}^{ (2) } = \lambda^{(2)} \big( 0, \, 1 \big)^T. $$
    \item Coupling to qubits $S_1$ and $S_2$: $$ \boldsymbol{\lambda}^{ (1,2) } = \lambda^{(1,2)} \big( 1, \, 1 \big)^T. $$
\end{enumerate}
If $d_1 = d_3$, the logical sub-spaces that generate each interaction pattern are given by
\begin{equation*}
\begin{gathered}
    \boldsymbol{c}^{ (1) } = \boldsymbol{F}^{-1} \cdot \boldsymbol{\lambda}^{ (1) } = \frac{d_1 (d_1 + d_2) \lambda^{ (1) }}{d_2 (2 d_1 + d_2) J}
    \begin{pmatrix}
    d_1 + d_2 \\ - d1
    \end{pmatrix}
    \\
    \boldsymbol{c}^{ (2) } = \boldsymbol{F}^{-1} \cdot \boldsymbol{\lambda}^{ (2) } = \frac{d_1 (d_1 + d_2) \lambda^{ (2) } }{d_2 (2 d_1 + d_2) J}
    \begin{pmatrix}
    - d1 \\ d_1 + d_2
    \end{pmatrix}
    \\
    \boldsymbol{c}^{ (1, 2) } = \boldsymbol{F}^{- 1} \cdot \boldsymbol{\lambda}^{ (1, 2) } = \frac{d_1 (d_1 + d_2) \lambda^{(1, 2)} }{ (2 d_1 + d_2) J}
    \begin{pmatrix}
    1 \\ 1
    \end{pmatrix}
\end{gathered}
\end{equation*}
Therefore, by setting $\max |c_i| = 1$, we find the maximum coupling for each case:
\begin{equation*}
\begin{gathered}
    \lambda^{ (1) }_{\max} = \frac{d_2 (2 d_1 + d_2 ) }{ d_1 (d_1 + d_2)^2 } J \\
    \lambda^{ (2) }_{\max} = \frac{d_2 (2 d_1 + d_2 ) }{ d_1 (d_1 + d_2)^2 } J \\
    \lambda^{ (1,2) }_{\max} = \frac{ 2 d_1 + d_2 }{ d_1 (d_1 + d_2) } J,
\end{gathered}
\end{equation*}
and if $d_1 = d_2 = d_3$ this simplifies to
\begin{equation*}
\begin{gathered}
    \lambda^{ (1) }_{\max} = \frac{3 J}{4 d} \\
    \lambda^{ (2) }_{\max} = \frac{3 J}{4 d} \\
    \lambda^{ (1, 2) }_{\max} = \frac{3 J}{2 d} .
\end{gathered}
\end{equation*}

\section{2D set-up: 8 ions}
\label{app:2d:ions}

Assume 8 ions trapped in a 2D lattice forming a cross as it is shown in Fig.~\ref{fig:SM:2D}, where the four inter ions constitute the control system and the external four ions correspond to the target system. Here we also consider a coupling between ions given by $f_{ij} = J |\boldsymbol{r}_i - \boldsymbol{r}_j |^{-1}$. Therefore, in this setting the adjacent qubits of the target system interact via the inherent ion-ion interaction with a coupling strength given by $f^S_{12} = J \sqrt{2} / ( d_1 + 2 d_2 ) $, meanwhile, the diametrically opposed ions interact with a coupling strength of $f^S_{13} = J / ( d_1 + 2 d_2 )$. The average interaction coupling of the inherent ion-ion in the target system is given by
\begin{equation*}
    \left\langle f^S_{ij} \right\rangle = \frac{1}{3} \, f^S_{13} + \frac{2}{3} \, f^S_{12} = \frac{1 + 2 \sqrt{2}}{3 d_1 + 6 d_2} J.
\end{equation*}

Due to the symmetry in the spatial distribution of the qubit of the control and the target system, each of the possible interaction patterns is equivalent (up to some relabeling of the qubits) to one of the following 5 cases:
\begin{enumerate}
    \item Coupling to all four qubits: $$\boldsymbol{\lambda}^{(1, 2, 3, 4)} = \lambda^{(1, 2, 3, 4)} \big(1, \, 1, \, 1, \, 1\big)^T.$$
    \item Coupling to qubit $S_1$: $$\boldsymbol{\lambda}^{(1)} = \lambda^{(1)} \big(1, \, 0, \, 0, \, 0\big)^T.$$
    \item Coupling to qubit $S_1$ and $S_2$: $$\boldsymbol{\lambda}^{(1,2)} = \lambda^{(1,2)} \big(1, \, 1, \, 0, \, 0\big)^T.$$
    \item Coupling to qubit $S_1$ and $S_3$: $$\boldsymbol{\lambda}^{(1,3)} = \lambda^{(1,3)} \big(1, \, 0, \, 1, \, 0\big)^T.$$
    \item Coupling to qubit $S_1$, $S_2$ and $S_3$: $$\boldsymbol{\lambda}^{(1,2)} = \lambda^{(1,2,3)} \big(1, \, 1, \, 1, \, 0\big)^T.$$
\end{enumerate}
The logical sub-spaces that generate them are given by $\boldsymbol{c} = \boldsymbol{F}^{-1} \cdot \boldsymbol{\lambda}$. If $d_1 = d_2 = d$, we obtain
\begin{equation*}
\label{eq:c:2dexample}
\begin{aligned}
    \boldsymbol{c}^{(1,2,3,4)} & = 0.36 \frac{d \lambda^{(1, 2, 3, 4)}}{J} \, \big( 1, \ 1, \, 1, \, 1 \big)^T \\
    \boldsymbol{c}^{(1)} & = \frac{2 d \lambda^{ (1) }}{13 J} \, \big( 14, \, - 2 \sqrt{10}, \, 1, \, - 2 \sqrt{10} \big)^T \\
    \boldsymbol{c}^{(1,2)} & = \frac{2 d \lambda^{(1, 2)}}{13 J} \, \big( 7.68 \, 7.68 \, -5.32 \, -5.32 \big)^T \\
    \boldsymbol{c}^{(1, 3)} & = \frac{2 d \lambda^{(1, 3)}}{13 J} \, \big( 15, \, -4 \sqrt{10}, \, 15, \, -4 \sqrt{10} \big)^T \\
    \boldsymbol{c}^{(1, 2, 3)} & = \frac{2 d \lambda^{(1, 2, 3)}}{13 J} \, \big(
    8.68, \, 1.35, \, 8.68, \, -11.65 \big)^T
\end{aligned}
\end{equation*}

Then the maximum coupling for each interaction pattern is given by
\begin{equation*}
\begin{gathered}
    \lambda_{\max}^{(1,2,3,4)} = 2.76 \frac{J}{d} \\
    \lambda_{\max}^{(1)} = 0.46 \frac{J}{d} \\
    \lambda_{\max}^{(1,2)} = 0.84 \frac{J}{d} \\
    \lambda_{\max}^{(1,3)} = 0.43 \frac{J}{d} \\
    \lambda_{\max}^{(1,2,3)} = 0.56 \frac{J}{d},
\end{gathered}
\end{equation*}
while the inherent interactions between the ions are given by
\begin{equation*}
\begin{gathered}
    f^S_{12} = 0.47 \frac{J}{d} \\
    f^S_{13} = 0.33 \frac{J}{d}.
\end{gathered}
\end{equation*}
With the values of $\lambda^{S'}$, we compute the time to implement all $U^{S'}$ what is given by $t = \pi/(4 \lambda^{S'})$:
\begin{equation*}
\begin{gathered}
    e^{-i \, \frac{\pi}{4} \, Z^C (Z_1^S + Z_2^S + Z_3^S + Z_4^S)} \quad \mapsto \quad t_{1234} = \frac{\pi d}{ 11.04 J} \\
    e^{-i \, \frac{\pi}{4} \, Z^C Z_1^S } \quad \mapsto \quad t_1 = \frac{\pi d}{ 1.84 J} \\
    e^{-i \, \frac{\pi}{4} \, Z^C ( Z_1^S + Z_2^S)} \quad \mapsto \quad t_{12} = \frac{\pi d}{3.36 J} \\
    e^{-i \, \frac{\pi}{4} \, Z^C ( Z_1^S + Z_2^S)} \quad \mapsto \quad t_{13} = \frac{\pi d}{1.72 J}. \\
    e^{-i \, \frac{\pi}{4} \, Z^C ( Z_1^S + Z_2^S + Z_3^S)} \quad \mapsto \quad t_{123} = \frac{\pi d}{2.24 J}. \\
\end{gathered}
\end{equation*}
Finally, we compute the time to implement a general logical gate in the control system:
\begin{equation}
    t_V = 2\left( \frac{\pi}{4 f_{12}^C} + \frac{\pi}{4 f_{23}^C} + \frac{\pi}{4 f_{34}^C} \right) = \frac{ 3\sqrt{2} d \pi}{J},
\end{equation}
where we consider the following circuit
\begin{equation*}
    V^C = \text{CX}^C_{3\to4}\text{CX}^C_{2\to3}\text{CX}^C_{1\to2} V^C_1 \text{CX}^C_{1\to2}\text{CX}^C_{2\to3}\text{CX}^C_{3\to4}.
\end{equation*}
\end{document}